\documentclass[aps,prd,showpacs,nofootinbib]{revtex4}
\usepackage[utf8]{inputenc}
\usepackage{subfigure}
\usepackage{verbatim}
\usepackage{graphicx}
\usepackage{mathrsfs}
\usepackage{color}
\usepackage{psfrag}
\usepackage{graphicx}
\usepackage{enumerate}
\usepackage{amsmath,amssymb,calc,amsfonts}
\usepackage{latexsym}
\usepackage{textpos}
\def\beq{\begin{eqnarray}}
\def\eeq{\end{eqnarray}}

\def\={\stackrel{\Delta}{=}}

\def\lie{\pounds}
\def\mcD{\mathcal{D}}
\def\mcS{\mathcal{S}}
\def\mcR{\mathcal{R}}

\DeclareMathOperator\erf{Erf}

\begin{document}
\title{Gravitational Collapse in the EGB Gravity}

\author{Ayan Chatterjee} \email{ayan.theory@gmail.com} 
\affiliation{Department of Physics and Astronomical Science,\\
Central University of Himachal Pradesh, Dharamshala-176206, India.}
\author{Avirup Ghosh} \email{avirup@ustc.edu.cn}
\affiliation{Interdisciplinary Center for Theoretical Study, 
University of Science and Technology of China, \\
Peng Huanwu Center for Fundamental Theory, Hefei, Anhui 230026, China }
\author{Suresh C. Jaryal}\email{suresh.fifthd@gmail.com}
\affiliation{Department of Physics and Astronomical Science,\\
Central University of Himachal Pradesh, Dharamshala-176206, India.}


\begin{abstract}
The Einstein- Gauss- Bonnet (EGB) gravity is an important modification of the Einstein 
theory of gravity and, for many gravitational phenomena, the Gauss- Bonnet (GB) correction 
term leads to drastic differences. In this paper, we study gravitational collapse in 
the $5$-dimensional EGB theory.
We construct the spherical marginally trapped surfaces and determine the evolution
of marginally trapped surfaces when the infalling matter admits
a wide variety of initial density distribution. We show that
the location of black hole horizon depends
crucially on the initial density and velocity profile of the
inflating matter as well as on the GB coupling constant.
A detailed comparison is made with the results of Einstein's theory.
\end{abstract}

\maketitle
\section{Introduction}
The study of gravitational collapse of a self- gravitating isolated system remains a
matter of great physical importance in understanding large scale structures in 
the universe, as well as towards discerning the formation of black hole horizons, 
spacetime singularities and the cosmic censorship
conjecture \cite{Hawking_Ellis, Wald, Landau_Lifshitz, joshi, Penrose:1964wq,Penrose:1969pc}.
In general relativity (GR), the spherical gravitational collapse and the singularity 
theorems have been studied at length. Although several important aspects including the 
cosmic censorship, non- symmetrical collapse remain to be understood completely, 
the progress in this direction has been remarkable.

The models of gravitational collapse in alternate theories of gravity,
including higher dimensional ones, 
are also being studied with interest 
since it is believed that one (or some) of these theories may
solve problems affecting GR, including spacetime singularities
\cite{Maeda:2006pm,Maeda:2007uu,Giambo:2007ps,Jhingan:2010zz,
Ghosh:2010jm, Taves:2011rd, Kunstatter:2012jr}. 
Among these, modified gravity
theories with higher curvature corrections arise naturally. 
Indeed, GR is viewed as an effective field theory
in which the Einstein- Hilbert action is only a low energy contribution
and higher curvature terms consistent with the diffeomorphism invariance may become
relevant as one goes to higher energies
\cite{Lanczos:1938sf, lanczos,Lovelock:1971yv,Lovelock:1972vz,
Buchbinder:1992rb,Biswas:2011ar,Caravelli:2010be,Modesto:2010uh}. 
Such higher curvature terms have been
explicitly obtained in string theories 
\cite{Deser:1974cz,Zwiebach:1985uq,Boulware:1985wk,Allemandi:2007bj,Nojiri:1999nd}. 
These higher curvature corrections 
should leave imprints at low energy scales which become important for low energy physics too,
affecting the horizon structure of large black holes. 
The Einstein- Gauss- Bonnet (EGB) theory is possibly 
the simplest diffeomorphism invariant modification of GR whose equations of 
motion contain no more than second order in time
derivatives \cite{lanczos,Lovelock:1971yv,Lovelock:1972vz, Zumino:1985dp,
ChoquetBruhat:1988dw,Deruelle:2003ck}. 
This generalization is also known to be the unique lowest 
order correction in the Lovelock action. Furthermore, since the EGB gravity is
free from ghosts (if the coupling constant has the same sign as the GR term) and leads 
to a well-defined initial value problem, it is
a respectable theory of gravity in higher dimensions, and its solutions have also been 
a matter of interest. In particular, black hole solutions
in the EGB theory are well known. They include the
Boulware- Deser, and other spherically symmetric solutions
\cite{Boulware:1985wk, Wheeler:1985nh, Wheeler:1985qd, Torii:2005xu}.
Black holes in EGB theory are also testbeds to gain fundamental insights
into various quantum aspects of gravity like the horizon entropy
 \cite{Jacobson:1993vj,Jacobson:1995uq,Chatterjee:2011wj}.

Thus, because of importance of the
EGB theory as a natural higher dimensional theory,
effect of the GB correction term on the spherically symmetric gravitational collapse 
and singularity structure have received attention. Naturally, particular
emphasis has been placed on the inhomogenous dust collapse models
of Lemaitre- Tolman- Bondi (LTB) type
\cite{Maeda:2006pm,Maeda:2007uu,Giambo:2007ps,Jhingan:2010zz,
Ghosh:2010jm}. In particular, \cite{Maeda:2006pm} has carried out a complete study of the
singularity structure of all the collapse models for spacetime dimensions $n\ge 5$.
It arises from this study that (i) all naked singularities for $n\ge 6$ are massless,
and (ii) for $n=5$, all singularities with mass $> 2\lambda $,  with $\lambda$ being the
GB coupling constant, are censored. This feature was also studied in the context of a
\emph{marginally bound} LTB spacetime by directly solving for the singularity curves,
and the apparent horizon, for a simple matter model \cite{Jhingan:2010zz}. Although,
some features of \cite{Maeda:2006pm} were borne out  in \cite{Jhingan:2010zz},
in particular that the central (as well as non-central) singularity is
naked, and this untrapped region increases with coupling constant $\lambda >0$,
it remains a possibility that these structures of local naked singularity
may well wash out if a more complicated or realistic matter profile is considered.
This expectation is not unwarranted since the occurrence of naked 
singularities break the Censorship conjecture \cite{Penrose:1969pc},
and the Seifert conjecture \cite{Seifert}, which essentially states that
massive singularities must be censored inside a trapped region. Of course, it remains a possibility
that these conjectures themselves need modifications in higher dimensions, just
like the Hoop conjecture \cite{Ida:2002hg}. Hence, it is 
essential that gravitational collapse in the $5$- dimensional EGB model be
studied in the full generality, using a large class of models where the matter admits
a wide variety of initial density and velocity profiles. This study shall, therefore
be useful to identify the region of the
parameter space where such singular structures arise.

Here, we develop the formalism of gravitational collapse in the EGB theory further
to the scenarios where, (a) the collapse is bounded (or, for that matter, unbounded), 
and that (b) density function of the collapsing matter 
has a realistic initial density distribution profile,
and (c) use this formalism to locate spherical (marginal) trapped surfaces
developing during the collapse of matter fields.
This shall be carried out by directly solving the equations of motion arising in the EGB theory
of gravity. This, to our knowledge, are significant improvements since
direct study using explicit solutions have been carried out
only for  marginally bound collapse models (see for example \cite{Jhingan:2010zz}).
Additionally, in the literature, the density of the collapsing 
matter profiles are restricted to simple power law models (including those carried out in \cite{Jhingan:2010zz}) and hence, 
these studies exclude possible realistic scenarios in which matter admits 
wider class of density distributions. Such density 
distributions include for example, a gaussian, or a matter profile with more complicated
dependence on space, including the angular coordinates (although, here we shall only concern ourselves
with matter profiles depending on radial coordinates).  
Indeed, the formation of spacetime singularity, the apparent horizon (AH),
the event horizon (EH) and their time development, depend not only
on the theory, or the initial velocity profile, but are also intimately connected with 
the density distribution of the collapsing matter. For example
in GR, the formation and dynamics of the AH changes drastically with variations
in the density profile \cite{Booth:2005ng, Chatterjee:2020khj}, and it is natural to expect that such time- development 
of horizons will also be observed for the GB modification too.

In this paper, we study these issues 
in the context of the inhomogeneous LTB collapse models in the EGB theory,
by carefully addressing them with examples. 
We track the motion of the collapsing shells, and simultaneously 
 follow the time development of horizon in relation to this collapsing matter.
In particular, we consider the 
horizon to be foliated by closed spherical $3$- dimensional
surfaces, such that the expansion scalar of the outgoing null normal vanishes $\theta_{(\ell)}=0$,
while that of the ingoing null normal is negative $\theta_{(n)}<0$. This
formulation of the black hole horizon is called Marginally Trapped Tube (MTT)
and has found use in analytical and numerical studies of 
black holes, in particular in understanding their classical nature, quantum behaviour,
as well as their stability under various geometric and physical variations \cite{Ashtekar:1997yu, Ashtekar:2000sz, Ashtekar:2000hw, Ashtekar:2002ag,Ashtekar:2003hk, Ashtekar:2004cn, Ashtekar:2005ez, Booth:2005ng, Andersson:2005gq,
Booth:2005qc, Booth:2006bn, Schnetter:2006yt,Andersson:2007fh,Chatterjee:2008if, Chatterjee:2012um,Chatterjee:2014jda,Chatterjee:2015fsa,Perez:2017cmj, Chatterjee:2020iuf}.
Note that since MTT is not associated with a particular signature, 
it can describe various states of a horizon.
For example, a black hole horizon in equilibrium
is a null MTT and is referred to as an isolated horizon (IH) (see \cite{Ashtekar:2000sz, Ashtekar:2000hw, 
Ashtekar:2004cn, Chatterjee:2008if,Chatterjee:2020iuf}.
A growing black hole admits a spacelike MTT, and is called
a dynamical horizon (DH) (see \cite{ Ashtekar:2002ag,Ashtekar:2003hk, Ashtekar:2004cn,
Chatterjee:2014jda,Chatterjee:2015fsa} for these horizons as well as their variations). Further, it is 
useful to describe a 
MTT with timelike signature, which admits matter flow in both directions,
and  is called a timelike tube. 
Thus MTTs provide an unified framework to study time evolution of black holes through different phases. 
The nature of spherical MTTs during gravitational collapse in GR 
has been studied in detail for various class of matter fields \cite{Chatterjee:2020khj, Booth:2005ng}. 
However, spherical MTTs in the EGB theory remains to be studied in the context of 
gravitational collapse of inhomogeneous matter fields (the LTB models), and here we fill this gap by 
making a detail study of these matter collapse models.
We carry out, (i) study the collapse end state with
special emphasis on the formation of horizons, and in particular, track the location of spherical marginally trapped tubes
with variation of matter profile, and (ii) for the mass profiles considered here, identify the
regions of the parameter space where the MTT evolves as a DH (spacelike),
where it might be timelike, and when it reaches equilibrium and become a null IH.
This shall also help us to  (iii) correctly locate the spherical outermost trapped surface 
developing during gravitational collapse. We must stress that although MTTs in $4$- dimensions
have been studied  \cite{Chatterjee:2020khj, Booth:2005ng}, their behaviour is drastically 
different in the EGB models, even for \emph{large} coupling constants.

The paper is arranged as follows: In the next section, we briefly discuss the equations of motions for the EGB theory
and it's reduction in the context of spherically symmetric spacetimes, in 
the $(t,\, r,\, \theta, \,\phi, \,\psi)$ coordinates. We shall also discuss the matter contributions to these
equations and the way to determine the spherically symmetric MTTs for these spacetimes. In section
\ref{section_3}, we solve the equations of motion directly for the marginally bounded and bounded cases.
The solution for the unbounded case is similar, and so we shall not repeat it here. We conclude in section IV
with discussions.


\section{Marginally trapped tubes in the EGB theory}
The formalism of MTT  as a quasilocal description of black hole horizons
was developed in \cite{Ashtekar:2005ez}. In the following, we present a brief 
discussion on this formalism, and set up the basic notations for our later use. 
Let us consider a $5$- dimensional spacetime $(\mathcal{M},g_{\mu\nu})$ with signature $(-,+,+,+, +)$.
Let $\Delta$ be a hypersurface  in $\mathcal{M}$
which may be spacelike,
timelike or even null.  
$\Delta$ is taken to be topologically $S^{3}\times \mathbb{R}$. 
At each point of the spacetime, we shall have 2 null vectors
and three spacelike vectors.
The null vector fields $\ell^{\mu}$ and $n^{\mu}$ are respectively 
the outgoing and the ingoing vector fields orthogonal to 
the $3$- sphere cross-sections of $\Delta$, with $\ell\cdot n=-1$.
The three normalised spacelike vectors tangential to the $3$- sphere are called
$\hat{\theta}$, $\hat{\phi}$, and $\hat{\psi}$ respectively,
and are orthogonal to the null vectors $\ell^{\mu}$ and $n^{\mu}$.
If $t^{\mu}$ is a vector field tangential 
to $\Delta$ and normal to the $S^{3}$ foliations,
then $t^{\mu}=\ell^{\mu}-Cn^{\nu}$. 
Now, assume that the $S^{3}$ foliations are such that
its null normals satisfy the following conditions:
(i) $\theta_{(\ell)}= 0$, and (ii) $\theta_{(n)} <0$.
The hypersurface $\Delta$ foliated by such surfaces 
is called a MTT. 
Note that MTT does not carry a specific signature.
Since $t\cdot t=2C$, the constant $C$ determines the signature of $\Delta$.
When $C=0$, $\Delta$ is null, foliated by $\ell^{\mu}$
and it describes a black hole in equilibrium (an IH).
it describes a black hole in equilibrium (an IH),
a DH when it is spacelike ($C>0$), or simply a timelike membrane when $C<0$
and $\Delta$ is timelike. Thus, MTT is an unified formalism for horizon evolution.
The value of $C$ can be 
determined for various gravitational collapse processes,
and for a wide class of energy momentum tensors.
Hence, the entire evolution of the MTT can be unambiguously
determined throughout the evolution process, if the
signature of $C$ is known.

As $t^{\mu}$ is orthogonal to the foliations 
and tangential to $\Delta$, it generates a foliation preserving flow so that 
on $\Delta$, the following condition holds:
\begin{equation}\label{liet}
\pounds_{t}\,\theta_{(\ell)}\triangleq 0.
\end{equation}
This equation implies that $C=\left[{\pounds_{\ell}\,\theta_{(\ell)}}/{\pounds_{n}\,\theta_{(\ell)}}\right]$. 
To determine the value of the constant $C$, we use the
geometrical equations of $3$-surface geometry given in the appendix \eqref{geometry_appendix}.
%
%
These equations imply that the constant $C$ which determines the nature of the MTT is given by:
\begin{equation}\label{value_of_c}
C=\frac{G_{\mu\nu}\,\ell^{\mu}\ell^{\nu}}{3(2\pi^{2}/\mathcal{A})^{2/3}- G_{\mu\nu}\,\ell^{\mu}n^{\nu}},
\end{equation}
where we have used the relation between area of the round $3$-sphere $\mathcal{A}$,
and the scalar curvature: $\mathcal{R}=6(2\pi^{2}/\mathcal{A})^{2/3}$.
We shall also assume that 
the Einstein- Gauss- Bonnet field
equations $G_{\,\mu \nu}\equiv R_{\mu \nu}-(1/2)R\,g_{\mu\nu}=T_{\,\mu\nu}$
\footnote{We use the 
units of $c=1$ and $8\pi G=1$, or equivalently, we scale the components of the 
energy- momentum tensor by $8\pi G$. In case
of the EGB theory too, we shall write the
Einstein equations in the similar manner, $G_{\mu\nu}=T_{\mu\nu}$.
In that case, $T_{\mu\nu}$ shall
imply a sum of terms, due to matter variables $T_{\mu\nu}$ and,
due to extra geometric variables arising out of the GB correction.}, holds on $\Delta$.

The signature of $C$ in eqn. \eqref{value_of_c}
is a quantity of utmost importance since it decides the nature and \emph{stability} of horizon
\cite{Andersson:2005gq,Andersson:2007fh}, and, as may be observed from the above equation, this value is regulated by 
the null components of the energy- momentum tensor as well as area of the cross-sections of the MTT.
However, in the following sections where we shall treat a wide class
of energy-momentum tensors for collapse models of the LTB type,
we shall observe that details like the initial velocity
profile, initial density profile of the collapsing matter, and the dimension of the spactime 
play important role as well. Indeed, in several cases, simple
changes in the density profile alters the nature and time of formation of 
the spacetime singularity, and that of the MTT quite drastically.
For example, in $4$- dimensions, if the matter profile is smooth, the MTT begins as a spacelike
hypersurface from the center of the cloud as soon as matter begins to fall, and asymptotes
to the null event horizons as infall of matter is discontinued.
Trapped surfaces in $4$- dimensions are discussed
in  \cite{Schnetter:2006yt,Booth:2005ng,Hajicek:1986hn, Wald:1991zz,Hayward:1993wb,krasinski_hellaby, Booth:2010eu,
Bengtsson:2008jr,Bengtsson:2010tj,
 Bengtsson:2013hla,Booth:2012rm, Creelman:2016laj,Booth:2017fob}.
However, in the $5$- dimensional EGB theory, even for the 
collapse of marginally bound matter with density admitting a Gaussian distribution,
the central singularity forms earlier than the corresponding MTT. This happens because
the EGB equations allow the formation of MTT only at the later shell coordinates, and hence,
the collapse of the first few shells leads to an untrapped singularity.    

In the following section, we shall discuss the EGB equations of motion for the 
spherical collapse of matter fields, and determine the requirements for
 formation of trapped surfaces in the $5$- dimensions.

\subsection{The equations of motion}
The action for the $5$- dimensional EGB theory is given by
\begin{eqnarray}
S&=&\int d^{5}x \sqrt{-g}\, (R+\lambda L_{GB})+S_{matter}, \label{actionEGB}
\end{eqnarray}
where $R$ is the Ricci scalar, $g$ denotes determinant of the metric $g_{\mu\nu}$ and, $\lambda$ is coupling constant of the Gauss- Bonnet term. The Gauss- Bonnet Lagrangian $(L_{GB})$ is given by
\begin{eqnarray}
L_{GB}&=& {R}^2-4 R_{\mu\nu} R^{\mu\nu}+R_{\mu\nu\sigma\delta}R^{\mu\nu\sigma\delta}. \label{GBLag}
\end{eqnarray}
The action eqn. \eqref{actionEGB} leads to the following field equations
\begin{eqnarray}
G_{\mu\nu}&\equiv&R_{\mu\nu}-\frac{1}{2}R\,g_{\mu\nu}=T_{\mu\nu}-\lambda H_{\mu\nu}\label{EGBQ},
\end{eqnarray}
where the term $G_{\mu\nu}$ is the usual Einstein tensor as in GR, $T_{\mu\nu}$ is the energy momentum tensor,
and $H_{\mu\nu}$ is the contribution due to the Gauss- Bonnet term. In the above equation \eqref{EGBQ}, the term $H_{\mu\nu}$ signifies the following 
\begin{eqnarray}
H_{\mu\nu}&=& H^{\prime}_{\mu\nu}-\frac{1}{2}g_{\mu\nu}\,L_{GB}\nonumber\\ 
&=&2 \left[R R_{\mu\nu}-2R_{\mu\lambda}R^{\lambda}\,_{\nu}-2R^{\lambda\sigma}R_{\mu\lambda\nu\sigma}+R_{\mu}\,^{\lambda\sigma\delta}\,R_{\nu\lambda\sigma\delta} \right]-\frac{1}{2}g_{\mu\nu}\,L_{GB}.\label{Hab}
\end{eqnarray}
Note that $H_{\mu\nu}$ may be considered as an effective energy momentum 
tensor adding to the usual matter tensor.

Now, we consider a general spherically symmetric collapsing cloud of fluid 
bounded by a spherical surface. 
In the comoving coordinates, the line element of 
a $5$ dimensional spherically symmetric spacetime geometry can be written as
\begin{equation}
 ds^{2}=-e^{2\alpha(r,t)}dt^2 + e^{2\beta(r,t)}dr^2 + R(r,t)^2 \left[d \theta^2+ \sin^2{\theta}\, d\phi^2+\sin^2{\theta}\,\sin^2{\phi}\, d\psi^2\right],
 \label{1eq1EGB}
\end{equation}
where $\alpha(r,t)$, $\beta(r,t)$ and $R(r,t)$\footnote{The symbol $R$ is used to denote both Ricci scalar and the radius of the matter configuration. We deliberately kept the same symbol since they will not appear simultaneously to cause any confusion. } are metric functions to be determined. $R(r,t)$ is radius of the collapsing matter cloud whereas, $\theta$, $\phi$, $\psi$ are the angular coordinates of that $3$-sphere. The energy momentum tensor for the fluid is taken to be 
\begin{eqnarray}
T_{\mu\nu}=(p_t+\rho)u_\mu u_\nu +p_t g_{\mu\nu}
+(p_r-p_t)X_{\mu}X_{\nu}\label{tmnzEGB}
\end{eqnarray}
where  $\rho(r,t)$ is density, whereas $p_{r}(r,t)$
and $p_{t}(r,t)$ are the radial and tangential components of pressure.
The $u^{\mu}$ and $X^{\mu}$ are unit 
time-like and space-like vectors satisfying $u_\mu u^\mu=-X_\mu X^\mu=-1$
In the comoving co-ordinates the four velocity and the unit space-like vector of the fluid as 
$u^\mu=e^{-\alpha}(\partial_{t})^{\mu}$ and $X^\mu=e^{-\beta}(\partial_{r})^{\mu}$.

The equation of motion for this metric in the EGB theory are given by
 \begin{eqnarray}
 \rho(r,t)&=&\frac{3}{2}\frac{F^{\,\prime}(r,t)}{R^{3}\, R^{\,\prime}},\label{1eq2EGB}\\
 {p}_{r}(r,t)&=&-\frac{3}{2}\frac{\dot{F}\,(r,t)}{R^2 \,\dot{R}},\label{1eq3EGB}\\
 \dot{R}\,{}^{\prime}&=&\dot{R}\,\alpha'+R\,{}^{\prime}\,\dot{\beta} ,  \label{1eq4EGB}\\
 \alpha^{\prime}&=&\frac{3R^{\, \prime}}{R}\frac{p_{t}-p_{r}}{\rho+p_{r}}- \frac{{p}^{\,\prime}_{r}}{\rho+p_r}  \label{1eq3egb}  ,\\
 F(r,t)&=&R^{2}\,(1-G+H)+ 2 \lambda (1-G+H)^2, \label{1eq5EGB}
 \end{eqnarray}
 where the superscripts primes (${}^{\prime}$) and dots ($\cdot$) represent partial
 derivatives with respect to $r$ and $t$ respectively.
 The quantity $R(r,t)$ is physical radius for matter configuration and $F(r,t)$ is 
 the Misner-Sharp mass function.
 The first and the second equations, \eqref{1eq2EGB} and \eqref{1eq3EGB}, are 
 the $G_{00}$ and the $G_{11}$ equations.
 The third is the $R_{01}$ equation. 
 The fourth equation is the Bianchi identity $\nabla_{\mu}T^{\mu r}=0$, which for the
 pressureless matter implies that the metric variable $\alpha^{\prime}=0$.
The equation \eqref{1eq5EGB},
 is the equation for the mass function with the functions $H(r,t)$ and $G(r,t)$  
defined as $H=e^{-2\alpha}\dot{R}^2$ and $G=e^{-2\beta}R^{\prime \,2}$. 

 Several points are to be noted regarding the abovementioned equations of motion.
 First, the relation between the matter variables and the geometric variables in 
 the above equations \eqref{1eq1EGB}- \eqref{1eq5EGB} are modified in comparison to 
 the $4$- dimensional Einstein theory. The changes in the numerical factors are due 
 to dimensionality of the spacetime as well as due to change in the theory itself,
 see for example equation \eqref{1eq5EGB}. 
 
Second, the number of independent
equations are five in number. 
The unknown functions in this problem are the three metric variables $\alpha(t,r)$, 
$\beta(t,r)$, $R(t,r)$, three matter variables $p_{r}(r,t)$, $p_{t}(t,r)$, $\rho(t,r)$,
and the mass-function $F(t,r)$. This combination allows two freely specifiable functions.
Since the equations give dynamical evolution of the functions, it is natural to 
specify these functions at an initial time $t=t_{i}$, and allow the
Einstein equations to evolve the dynamical functions. 
Since we shall be dealing with pressureless (dust) collapse,
it is useful to point out that for dust collapse, $p_{r}$ and $p_{t}$ 
are taken to vanish at $t_{i}$, and this fixes the function $\alpha(t,r)=\alpha(t)$.
We shall show below that this effectively implies $\alpha=0$, since we can rescale
the time coordinate. 
The remaining freely specifiable functions are the density $\rho(r,t_{i})$, and $\beta(r,t_{i})$
which, as we shall show below, implies the specification of initial density and velocity profiles of 
the collapsing matter. We shall also assume that $R(r,t_{i})=r$. This requirement is
consistent with the regularity conditions discussed below. By choosing different values of 
$r$ at the initial surface gives the time evolution of the various shells of matter.  

Thirdly, few regularity conditions on the metric functions must also be enforced during 
the collapse process.
The positivity and regularity of the density $\rho(t,r)$, and equation \eqref{1eq2EGB} imply that
the mass function $F(t,r)$ must smoothly vanish at the center of 
the matter configuration at $r=0$. The condition $R(t,r)=0$ is the genuine spacetime singularity
where the density and the curvature scalars blow up.
Note that the density also
blows up for $R^{\,\prime}=0$, although this is not a genuine spacetime singularity and can be removed.
This condition in fact implies shell-crossings,
when shells of fixed $r$ cross each other.
The sufficient condition which 
guarantees no shell-crossing is $R^{\,\prime}> 0$,
which ensures that shells maintain their ordering.  
Note that $R(r,t_{i})=r$, and any other index on $r$ leads either to shell- crossing 
or affects differentiability of metric functions at center of the matter configuration. 
An important requirement for gravitational collapse is 
to require $\dot{R}(r,t)<\,0$. Finally, we shall ensure
in our study that no trapped surface is present at the initial data, by checking that 
the value of $r$ at the initial surface is greater than the condition of formation of trapped
surface at that coordinate.


Now, with the metric given in equation \eqref{1eq1EGB}, the outgoing and the incoming 
null normals to the $3$-sphere are given by\footnote{These expressions
are valid for dust collapse. In general, one has $e^{-\alpha(r,t)}\, (\partial_{\,t})^{\mu}$ 
in place of $(\partial_{\,t})^{\mu}$ in 
equations \eqref{null_normals_exp} and \eqref{null_normals_exp_2}.}:
\begin{eqnarray}\label{null_normals_exp}
\ell^{\mu}&=&(\partial_{\,t})^{\mu}+ e^{-\beta(t,r)}\,(\partial_{r})^{\mu}\\
n^{\mu}&=&(1/2)(\partial_{\, t})^{\mu} - (1/2)\,e^{-\beta(t,r)}\,(\partial_{r})^{\mu}.\label{null_normals_exp_2}
\end{eqnarray}
This leads to the following expressions for the expansion scalars:
\begin{eqnarray}\label{theta_ell}
\theta_{(\ell)}&=& \frac{3}{R(r,t)} [\dot{R} + R^{\prime} \exp(-\beta)]
= \frac{3}{R(r,t)}[\dot{R} + \sqrt{1-k(r)}],\\
\theta_{(n)}&=& \frac{3}{R(r,t)}[\dot{R}- \sqrt{1-k(r)}],
\end{eqnarray}
where we have used the relation $ R^{\prime}= e^{\beta(r, t)}\,\sqrt{1-k(r)}$. 
This relation is obtained as follows: For 
the case of pressureless matter, eqn. \eqref{1eq3egb} gives $\alpha^{\prime}=0$
which along with eqn. (\ref{1eq4EGB}) implies:
\begin{equation}\label{gsoln}
G(t,r)=e^{-2\beta(r, t)}R^{\, \prime\, 2}\equiv E(r)\equiv 1-k(r),
\end{equation}
where $k(r), \,E(r)$ are the integration functions.
From the equation \eqref{1eq5EGB}, the equation of motion of collapsing configuration gives 
the following expression for $\dot{R}(t,r)$:
\begin{eqnarray}
\dot{R}(r,t)&=&-\left[(1/4\lambda)\{\sqrt{R^{4}+8\lambda F(r,t)} -R^{2}\} -k(r)\right]^{1/2}
\label{rdoteqn},
\end{eqnarray}
where we have used the $-$ve sign, as required for gravitational collapse. 
It follows from this equation \eqref{rdoteqn}, and the equation \eqref{theta_ell} that 
the condition for $\theta_{(\ell)}=0$ requires:
\begin{equation}\label{trapped_surface_eqn}
R_{M}(r,t)=\sqrt{F(r,t)-2\lambda},
\end{equation}
which at the same time is also the condition for $\theta_{(n)}<0$. 
Thus, for the spacetimes we are studying, all the three spheres which satisfy eqn.
\eqref{trapped_surface_eqn} are marginally trapped spheres.

As discussed earlier following eqn \eqref{value_of_c},
the dynamics of the marginally trapped surfaces 
(whether they are timelike, spacelike or null), depends upon sign of the expansion parameter $C$. On 
the trapped surface, it is defined by
\begin{eqnarray}
C=\frac{T_{\mu\nu}\ell^{\mu} \ell^{\nu}-\lambda H_{\mu\nu}\ell^{\mu} \ell^{\nu}}
{3/R(r,t)^2-T_{\mu\nu}\ell^{\mu} n^{\nu}+\lambda H_{\mu\nu}\ell^{\mu} n^{\nu}},\label{OSDCEGB}
\end{eqnarray}
where eqn.\eqref{value_of_c} and eqn.\eqref{EGBQ} have been used.
Now, the task is to write down all the components
in $T_{\mu\nu}$ as well as in $H_{\mu\nu}$
in terms of the matter variables. The expression for
$T_{\mu\nu}$ is already given in equation \eqref{tmnzEGB}.
The details of the calculation for $H_{\mu\nu}$ is carried out in the appendix \eqref{matter_appendix}.
The quantity $H_{\mu\nu}\,\ell^{\mu}\ell^{\nu}$ in equation \eqref{OSDCEGB} is given by:
\begin{eqnarray}
H_{\mu\nu}\ell^{\mu}\ell^{\nu}=2\left[\frac{{6F}\left(\rho+p_{r}\right)}{\left(F-2\lambda\right)^2}+2p_{\theta}^2-4p_{r}p_{\theta}-\frac{2}{3}p_{\theta}\left(\rho+p_{r} \right) \right],
\label{Hlalb}
\end{eqnarray}
Similarly, the expressions for $H_{\mu\nu}\,\ell^{\mu}\, n^{\nu}$ involves 
two terms, which are given by:
\begin{eqnarray}
H^{\prime}_{\mu\nu}\,\ell^{\mu}\,n^{\nu}
&=&2\left[4p_{\theta}\left(p_{\theta}+\frac{2}{9}\rho-\frac{4}{3}p_{r}\right)-\frac{2}{\left(F-2\lambda\right)^2}\{6F p_{t}+\left(F+4\lambda\right)\left(\rho-p_{r}\right)\}\right.\nonumber\\
&-&\left.6\left\{p_{t}+\frac{2}{3}\left(\rho-p_{r}\right)
-\frac{3F}{\left(F-2\lambda\right)^2} \right\}^{2}+\frac{16}{9}\left(\rho^2+p_{r}^2\right)-72\frac{\lambda^{2}}{\left(F-2\lambda\right)^{4}}\right].\nonumber\\ \label{Hlanb}
\end{eqnarray}
The term involving the $L_{GB}$ gives the following expression in terms of the matter variables:
\begin{eqnarray}
L_{GB}&=&R^{2}-4R_{tt}R^{tt}-4R_{rr}R^{rr}-12R_{\theta\theta}R^{\theta\theta}
+6R_{trtr}R^{trtr}+18R_{t\theta t\theta}R^{t\theta t\theta}
+18R_{r\theta r\theta}R^{r\theta r\theta}+18R_{\theta \phi\theta\phi}R^{\theta \phi\theta\phi}\nonumber\\
&=&\left[\frac{2}{3}\left(\rho-p_{r} \right)-2p_{t} \right]^2 
+ 18\left[ \frac{F^{2}+32\lambda^2}{\left(F-2\lambda\right)^{4}}\right]+6\left[p_{t}+\frac{2}{3}\left(\rho-p_{r}\right)-\frac{3F}{\left(F-2\lambda\right)^2} \right]^{2}\nonumber\\
&&~~~~~~~~ ~~~~~~~~~~~~~ ~~~~~~~~~~~
-\frac{12}{9}\left(\rho-p_{r} \right)^2-4 \left[\frac{2}{3}\left(\rho+p_{r} \right)+p_{t} \right]^2 -4 \left[\frac{2}{3}\left(\rho+p_{r} \right)-p_{t} \right]^2 .\label{LGBlanb}
\end{eqnarray}
Using these expressions in equation \eqref{OSDCEGB}, we shall understand 
the evolution of spherical MTTs for various collapse scenarios.

\section{Gravitational Collapse  for Pressureless Matter}\label{section_3}
%
Let us use the equations derived above to understand the 
dynamics of collapse process for pressureless matter configuration.
In the absence of pressure, 
the EGB equation \eqref{1eq3EGB}
implies that $F=F(r)$, whereas eqn. \eqref{1eq3egb} gives $\alpha^{\prime}=0$. 
The metric function $\alpha(t,r)$
is a function of $t$ only. This allows the rescaling of the time coordinate so that effectively
$\alpha(t,r)=0$. The metric function $\beta(t,r)$ follows from eqn. \eqref{gsoln}. This two
solutions implies that the metric is given by:
\begin{eqnarray}\label{int_metric}
ds^{2}=-dt^2 + \frac{R^{\, \prime \, 2}}{1-k(r)}\,dr^2 + R(r,t)^{\, 2} \,d\Omega_{3},
\end{eqnarray}
where $d\Omega_{3}$ is the metric of an
unit round $3$-sphere, and $R(t,r)$ is obtained from 
the equation \eqref{1eq5EGB}, which gives the equation of motion of 
the collapsing matter configuration in $5$D-EGB theory:
\begin{eqnarray}
\dot{R}^{\,2}(r,t)&=&-k(r)-\frac{R^2}{4\lambda}+\frac{R^2}{4\lambda}\left[1+\frac{8\lambda F}{R^4}\right]^{1/2},\label{EQMEGB}
\end{eqnarray}
where we have used eqn.\eqref{gsoln}. The function $k(r)$ can take either
signatures or zero. The situation where $k(r)$ remains vanishing during
the collapse process is called a marginally bound collapse, whereas the one in
which $k(r)$ admits a positive signature is called a bounded collapse. We shall 
deal with these two cases only. The behaviour for unbounded 
gravitational collapse in EGB theory is similar and shall not be carried out here.

Now, one has to ensure that this metric existing inside the
collapsing matter cloud must be matched to an exterior static spherically symmetric metric.
Such a metric is already well known as the Boulware- Deser- Wheeler 
solution \cite{Boulware:1985wk, Wheeler:1985nh, Wheeler:1985qd, Torii:2005xu}.
We shall always ensure that metric of the collapsing matter cloud remains matched to an external 
Boulware- Deser- Wheeler solution of mass $M$, across a timelike hypersurface $r_{b}$. As we show in
the appendix \eqref{matching_appendix}, such a matching leads to the condition that $F(r_{b})= M$.  
 
In the following, we shall consider a wide variety of density profiles
for matter fields and note the formation of singularity
and spherically symmetric trapped surfaces and horizons.

\subsection{Marginally bound collapse}
%
%
For the marginally bound collapse, we have $k=0$.
From the equations (\ref{1eq5EGB}) and \eqref{EQMEGB}, the equation of motion is
\begin{eqnarray}
\dot{R}\,^{2}(r,t)&=&-\frac{R^2}{4\lambda}+\frac{R^2}{4\lambda}\left[1+\frac{8\lambda F}{R^4}\right]^{1/2} \label{dotRk0EGB}.
\end{eqnarray}
Using some simple substitutions and algebra we get the equation for matter shells corresponding 
to values of $R(r,t)$ (see also \cite{Jhingan:2010zz})
\begin{eqnarray}
t_{sh}&=&t_{s}-\left[ \frac{\lambda R^2}{\sqrt{R^4-8\lambda F}-R^2}\right]^{1/2}
-\frac{\sqrt{\lambda}}{2\sqrt{2}}\tan^{-1}\left[\frac{3R^{2}-\sqrt{R^4-8\lambda F}}{2\sqrt{2}\left\{\sqrt{R^4-8\lambda F}-R^2\right\}^{1/2}} \right],
\label{tCollapsek0EGB}
\end{eqnarray}
where $t_s$ is the time of the formation of singularity, and is given by:
\begin{eqnarray}
t_{s}&=&\frac{\sqrt{\lambda}}{2\sqrt{2}}\tan^{-1}\left[\frac{3r^{2}-\sqrt{r^4-8\lambda F}}{2\sqrt{2}\left\{\sqrt{r^4-8\lambda F}-r^2\right\}^{1/2}} \right]+\left[ \frac{\lambda r^2}{\sqrt{r^4-8\lambda F}-r^2}\right]^{1/2}.\label{tsEGB}
\end{eqnarray}
The expression of the time for shells reach the Boulware- Deser-Wheeler horizon or the MTT,
obtained for $R(r,t)=\sqrt{F(r,t)-2\lambda}$ is 
given by $t_{AH}$:
\begin{eqnarray}
t_{AH}&=&t_{s}-\left[ \frac{\lambda \left( F-2\lambda\right)}{\sqrt{\left( F-2\lambda\right)^{2}-8\lambda F}-\left( F-2\lambda\right)}\right]^{1/2}-\frac{\sqrt{\lambda}}{2\sqrt{2}}\tan^{-1}\left[\frac{3 \left( F-2\lambda\right)-\sqrt{\left( F-2\lambda\right)^{2}-8\lambda F}}{2\sqrt{2}\left\{\sqrt{\left( F-2\lambda\right)^{2}-8\lambda F}-\left( F-2\lambda\right)\right\}^{1/2}} \right].
\label{tappk=0EGB}
\end{eqnarray}
Given these expressions we now proceeds to understand the nature of MTTs for some realistic mass profiles.
  
 %
\begin{figure}[h!]
\centering
\subfigure[\ ] {
\includegraphics[scale=0.25]{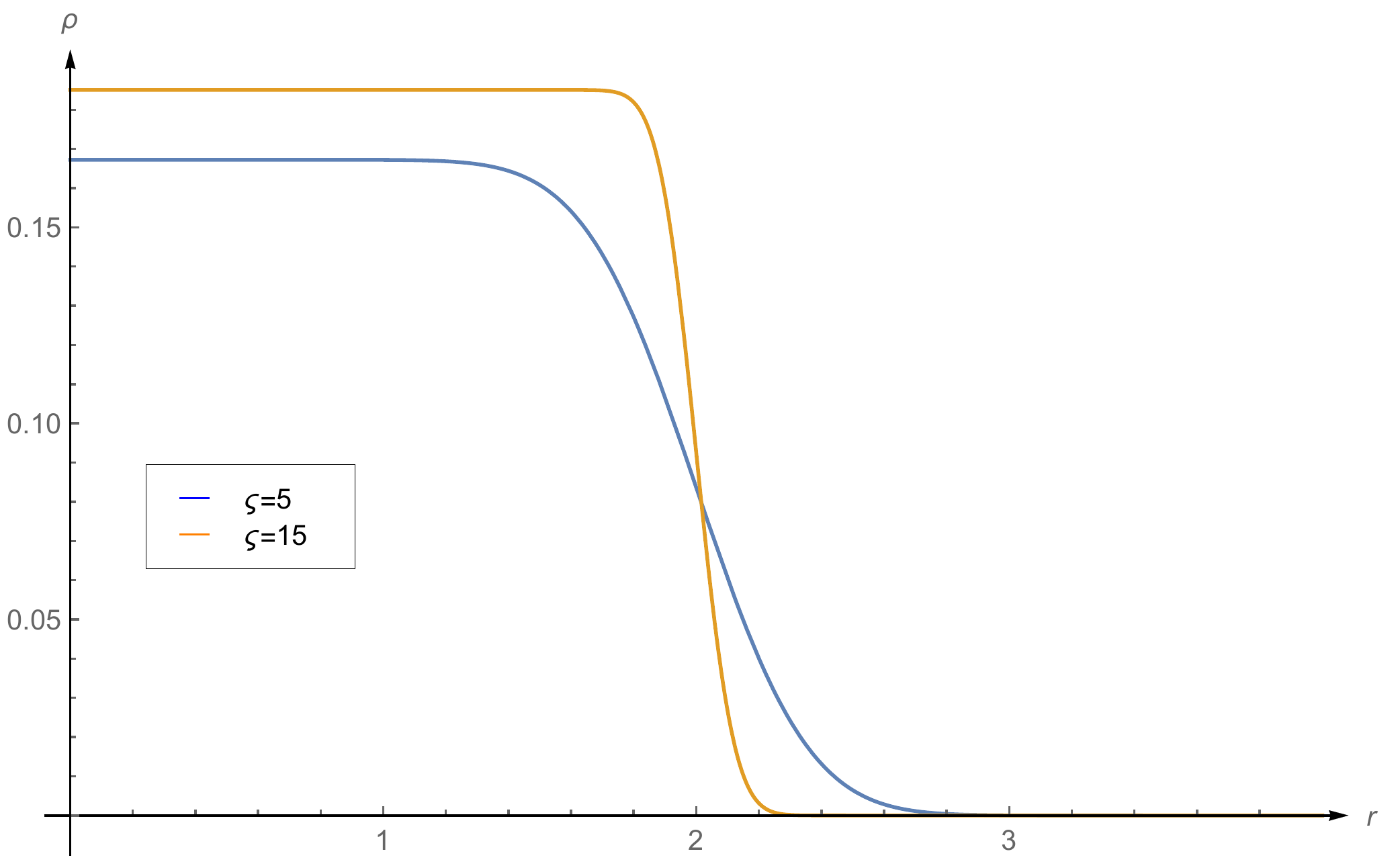}}
\subfigure[\ ] {
\includegraphics[scale=0.35]{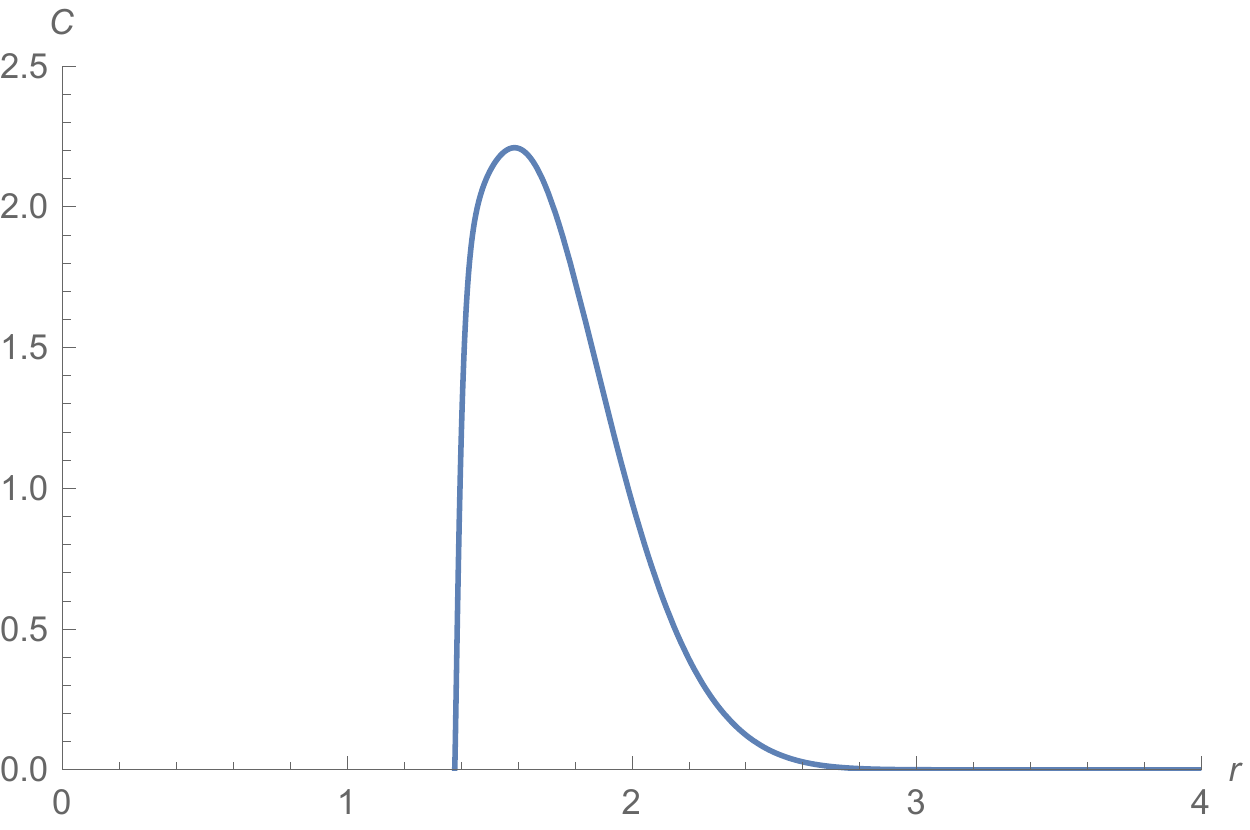}}
\subfigure[\ ] {
\includegraphics[scale=0.4]{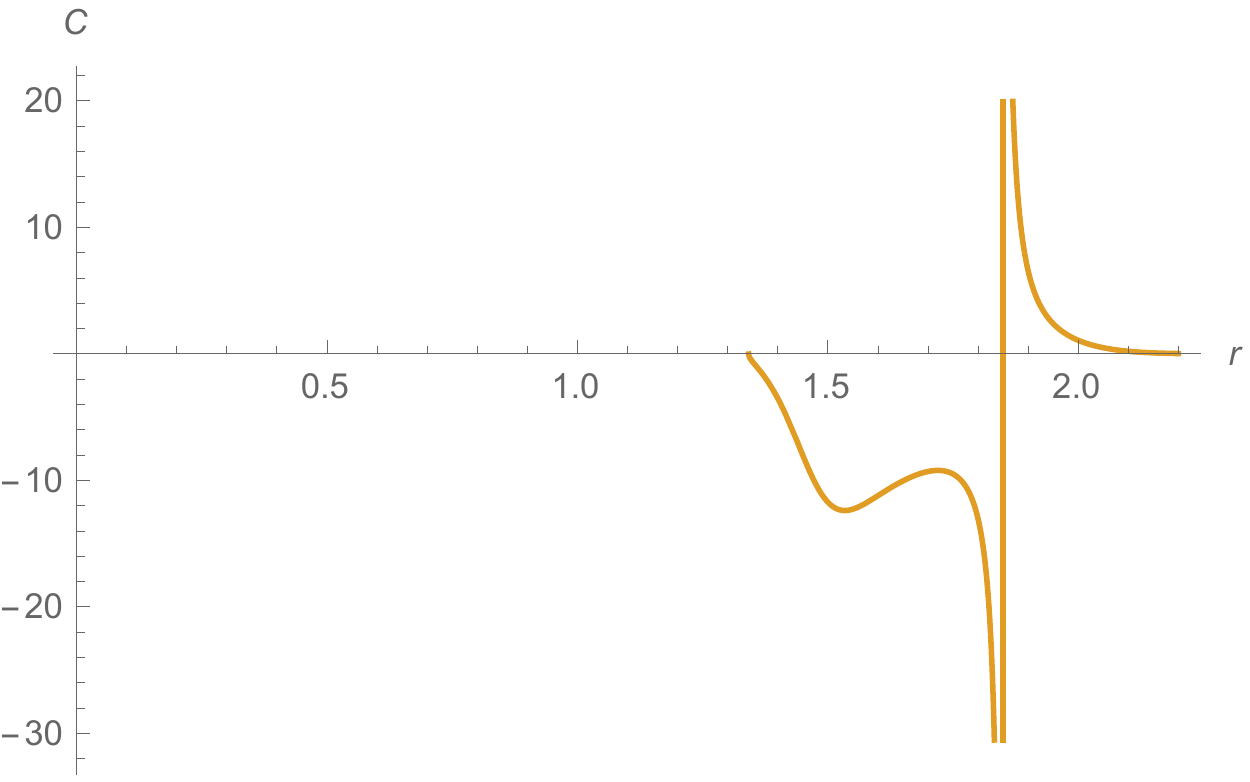}}
\subfigure[\ ] {
\includegraphics[scale=0.35]{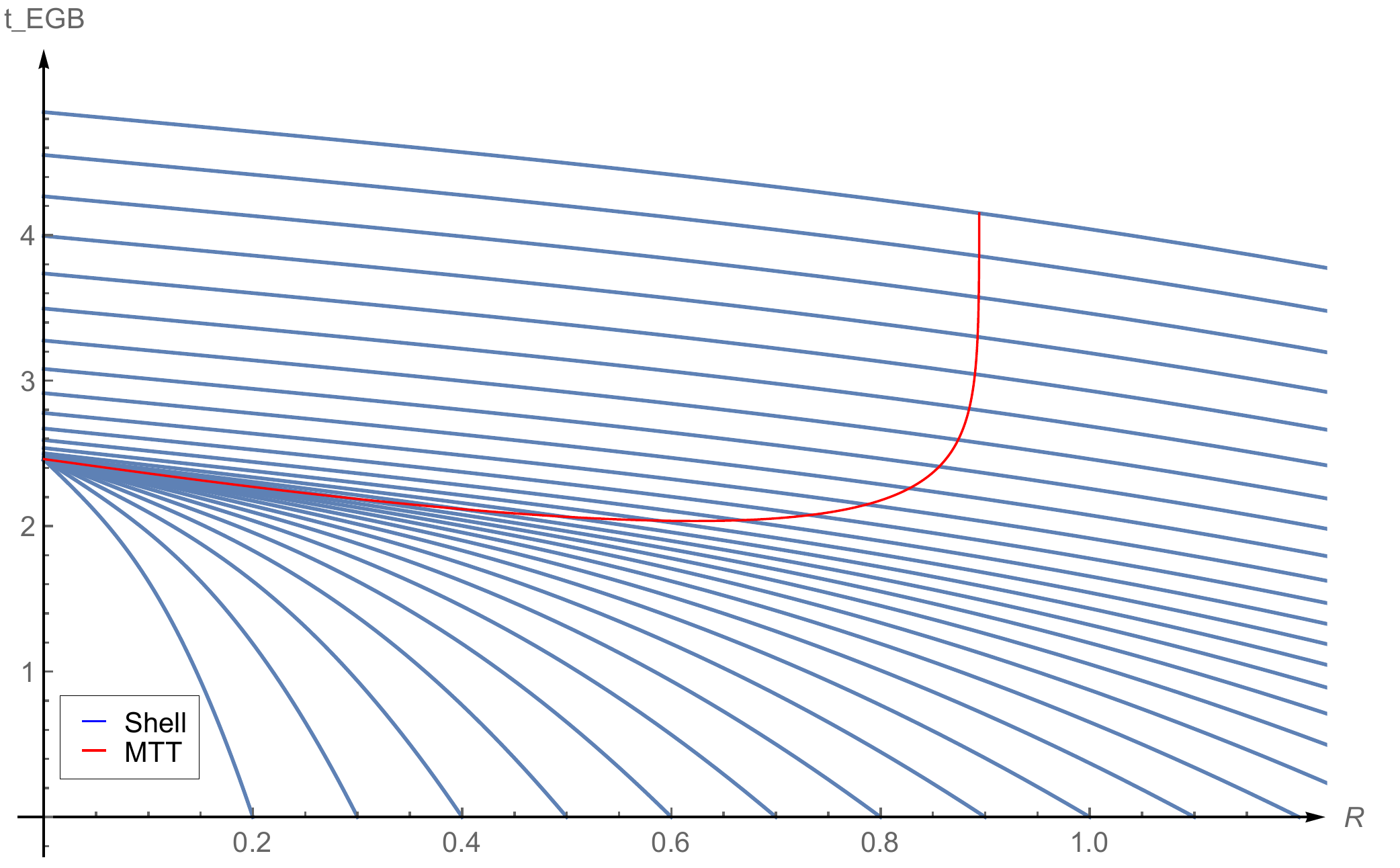}}
\subfigure[\ ] {
\includegraphics[scale=0.35]{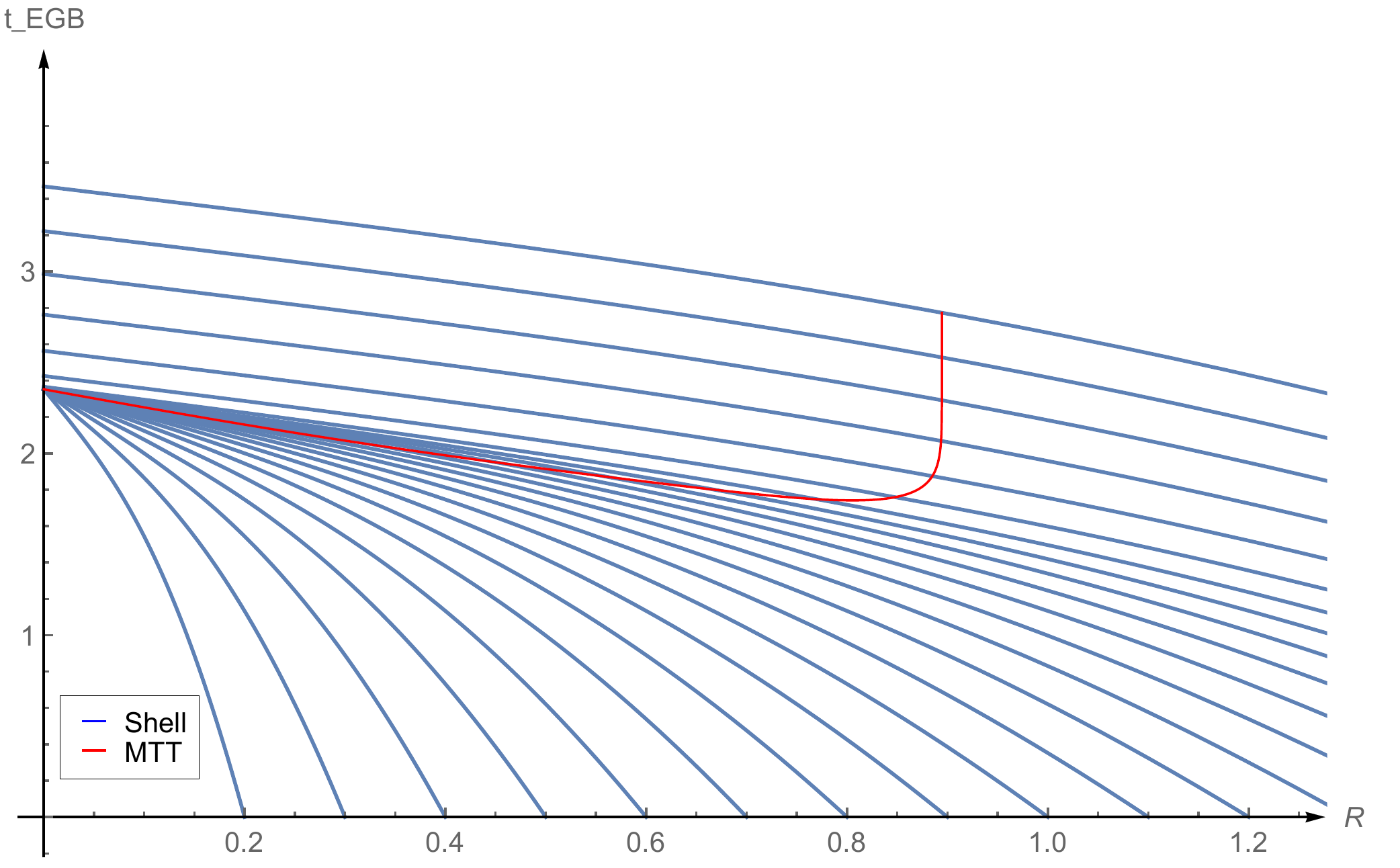}}
\caption{\label{fig:OSD_example1}
These figures give the gravitational collapse for the
density profile of eqn. \eqref{OSD_rho_dist}. 
For the plot we have used the EGB coupling constant $\lambda=0.1$. The
figure (a) gives the density fall-off for two choices of the control parameter $\varsigma$, (b) is the plot of the function $C$
for $\varsigma= 5$. The signature of $C$ shows that the MTT in this case is spacelike.
(c) is the plot of the function $C$
for $\varsigma=15$. The signature of $C$ shows that the MTT in this case is timelike.
The figure (d) is $R(r,t)$ vs $t$
graph for $\varsigma=5$, (e) gives the time development of MTT for
 $\varsigma=15$ along with the collapse of each shell. In the $R-t$ graphs, the shells are denoted by blue lines
 whereas the red lines are the MTT.  The straight vertical red lines in (d) and (e) 
represents the isolated horizon phase of the MTT and is reached when no more matter falls in.}
\end{figure}
%


%
 \subsubsection*{Examples} 
 \begin{enumerate}
 \item Let us first consider a collapsing matter profile
 which admits a variation in the density  
 distribution according to the choice of two parameters $\varsigma$
 and $r_{0}$.
 The density distribution is of the following form: 
 \begin{equation}\label{OSD_rho_dist}
 \rho(r)=\frac{m_{0}\mathcal{E}(\varsigma)}{r_{0}^{4}}\left[1-\erf\left\{\varsigma\left(\frac{r}{r_{0}}
 -1\right)\right\}\right],
 \end{equation}
 where $m_{0}=m(r\rightarrow \infty)$ is the total mass of the cloud,
 $r_{0}$ is the label on the matter shell coordinate where the variation
 of the density with the radial coordinate is largest, \emph{i.e} $-(d\rho/dr)$
 is highest. We shall choose the value of $r_{0}=2$.
 The parameter $\varsigma$ in equation \eqref{OSD_rho_dist} controls the variation
 of density function. A similar density profile was also studied for LTB models in $4$-d 
 GR \cite{Booth:2005ng, Chatterjee:2020khj}. 
 As seen from the plot in figure \eqref{fig:OSD_example1}(a), a larger 
 value of $\varsigma$ implies a 
 step- function- type distribution of the density, whereas,
 for a lower value of $\varsigma$, the density varies slowly with $r$. 
 So, $\varsigma$ is a control parameter for
  approach towards the OSD model- larger the value of $\varsigma$, closer is 
  the density to isotropy, and smaller values of $\varsigma$ implies inhomogeneities. 
  The function $\mathcal{E}(\sigma)$
  has the following form:
  \begin{equation}
  \mathcal{E}(\varsigma)=3\varsigma^{3}\, \left[2\pi\varsigma(2\varsigma^{2}+3)(1+\erf \varsigma)
  +4\sqrt{\pi}\exp(-\varsigma)(1+\varsigma^{2})\right]^{-1},
  \end{equation}
  and $\erf$ is the usual error function.
  We consider the cases where $\varsigma=5$ and $15$. The graphs are 
  given in figure \eqref{fig:OSD_example1}.
  
   From figure \eqref{fig:OSD_example1}(d), we note that as shells begin to collapse, 
   the MTT begins to form, and grows with the fall
   of the shells, until the growth stops when all the shells upto $r=2$ has fallen in.
   This happens since the matter density is almost zero after $r=2$. After all 
   the matter goes in, the MTT becomes null, as seen by the straight line in figure 
   \eqref{fig:OSD_example1}(d). 
   The MTT becomes null at $R= 0.89$ since the total mass of the cloud is unity, and 
   hence for $\lambda=0.1$, the MTT is obtained from eqn. \eqref{trapped_surface_eqn} 
   to be $\sqrt{0.8}=0.894$.In this region, the MTT
   has reached the IH phase. 
   
   Two further points need to be noticed. First, for $\varsigma=5$, the MTT
   are spacelike. This may be seen from the values of $C$ in figure \eqref{fig:OSD_example1}(b).
   However, if we look at the $R(r,t)-t$ graph in figure \eqref{fig:OSD_example1}(d),
   it seems that the MTT may have become timelike in certain regions. This
   apparent contradiction was also noted earlier in \cite{Andersson:2005gq, Booth:2005ng} 
   and happens due to non- trivial ways in which the MTT crosses the chosen folations.
    For $\varsigma=15$, the MTT is surely timelike, as may be noted from figures \eqref{fig:OSD_example1}(c) and 
    \eqref{fig:OSD_example1}(e). The MTT begins to form earlier at $r=1.7$ at $t=1.8$
    and then begins to grow on either side to match with the MTT at the center $R=0$,
    and also towards the IH at $R=0.89$. This possible points towards an 
    unstable MTT. as was pointed out in the case of GR in \cite{Andersson:2005gq, 
    Booth:2005ng,Chatterjee:2020khj}
    
    Secondly, as can be noted from the graph in figure \eqref{fig:OSD_example1}(e),
    all the shells, denoted by the blue lines reach the singularity at $R=0$ at the same time,
    which is a distinctive feature of the OSD process. As the value of $\varsigma$
    is lowered, the example of  \eqref{fig:OSD_example1}(d) shows that the shells begin to deviate marginally from
    this feature since the deviation in the density profile remains small.
    This also points to the fact that this collapse process is similar to that in GR, at least in this particular case of
    isotropic collapse.

\item For the next example, we take the 
mass density to have following form \cite{Jhingan:2010zz, Chatterjee:2020khj}:
\begin{eqnarray}
&&\rho(r)=m_{0}[1-(r/r_{0})]\,\Theta(100-r) \label{rhoLTBEGB}
\end{eqnarray}
where $\Theta(x)$ denotes the Heaviside theta function, and $r_{0}=100\, m_{0}$.
The graph of $\rho$, $C$ and $R(r,t)$-$t$ are given in 
the figure \eqref{fig:ltbk0_example2EGB}(a), \eqref{fig:ltbk0_example2EGB}(b),
and \eqref{fig:ltbk0_example2EGB}(c) respectively. 
Note that the MTT begin around $t=2900$ when the shell at $r=50$
has already fallen in. After this growth, it remains a dynamical horizon 
throughout and becomes an isolated horizon only when the matter shells stops falling at $r=100$
and all the matter has collapsed. This behaviour in the $R-\, t$ plot is
reflected in the graph of $C$ quite faithfully. Indeed, the signature of $C$ indicates 
that MTT is spacelike, beginning at $r=50$ and continues until 
the shell at $r=100$ falls, after which it becomes null.
 
Note however that MTT does not begin to form immediately,
but only after some shells have fallen in. This is because of a simple reason
but leads to some important consequences, and is discussed below:
The MTT forms only when the condition in eqn. \eqref{trapped_surface_eqn} is satisfied.
Indeed, for the early shells, the value of $F(r)$ for these shells, \emph{i.e.}
the amount of matter contained inside the sphere of radius $r$ at the initial time,
is smaller than the value of $\lambda$, which here
is taken to be $0.1$. For that reason, $R_{M}(r,t)$ does not admit real values.
It is only after sufficient number of shells have fallen in, that 
condition of trapped surface can be evaluated to obtain a real value. 
Until that time, the central singularity
remains naked for a trapped surface. Our study reveals 
this feature in a direct manner since we 
have been able to probe each and every matter shells quite elaborately. 

%
\begin{figure}[t!]
\centering
\subfigure[\ ] {
\includegraphics[scale=0.4]{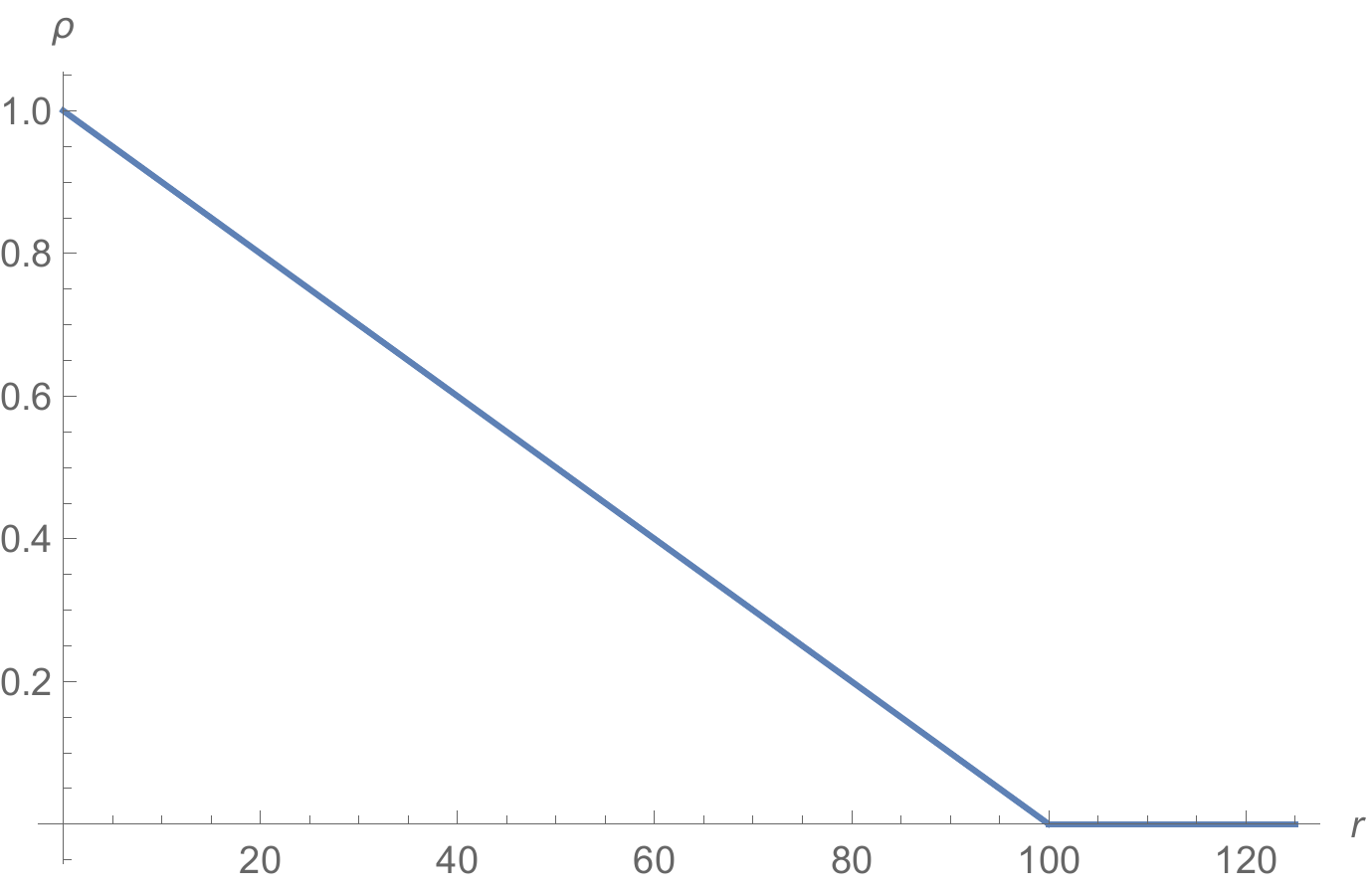}}
\subfigure[\ ] {
\includegraphics[scale=0.4]{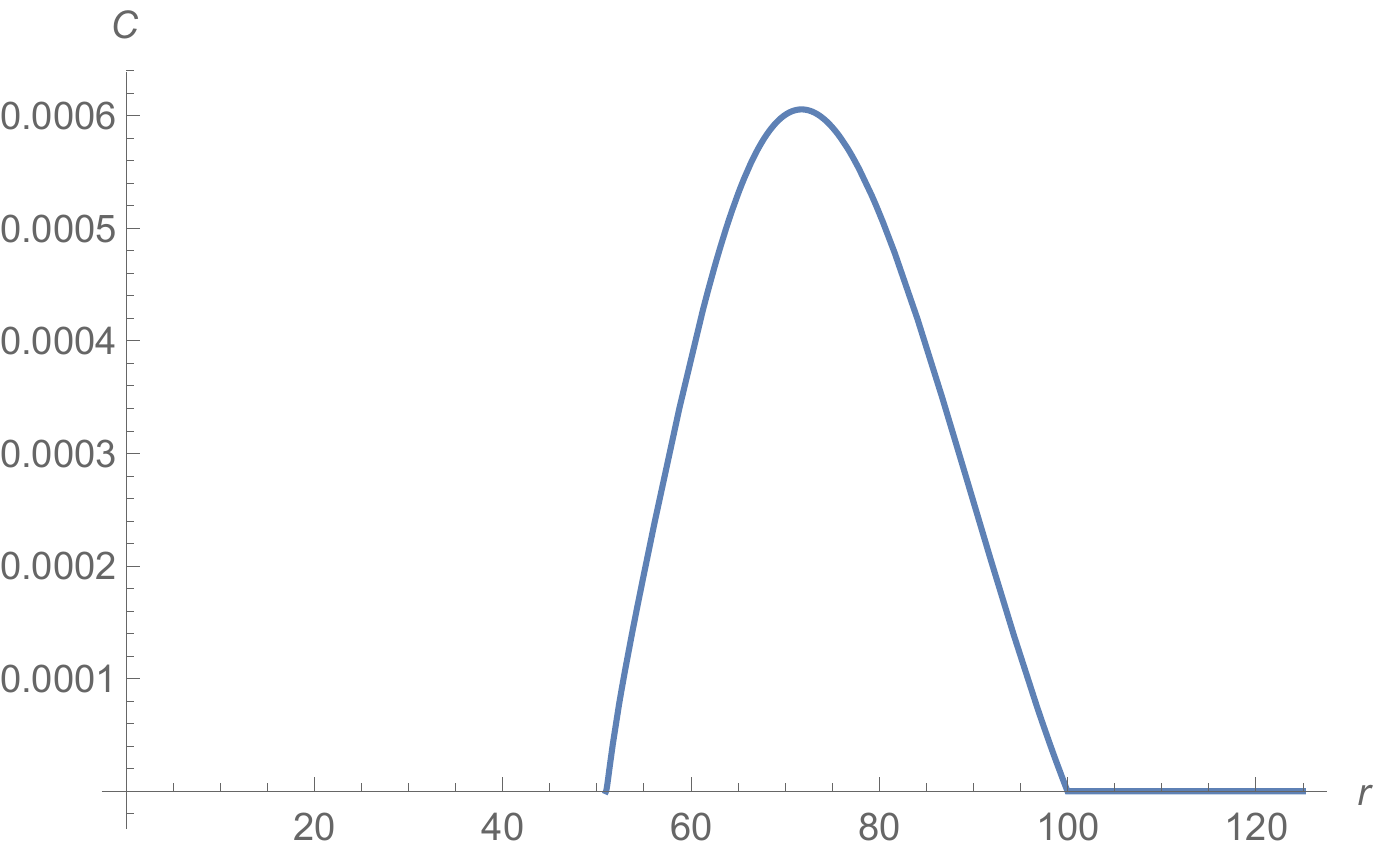}}
\subfigure[\ ] {
\includegraphics[scale=0.4]{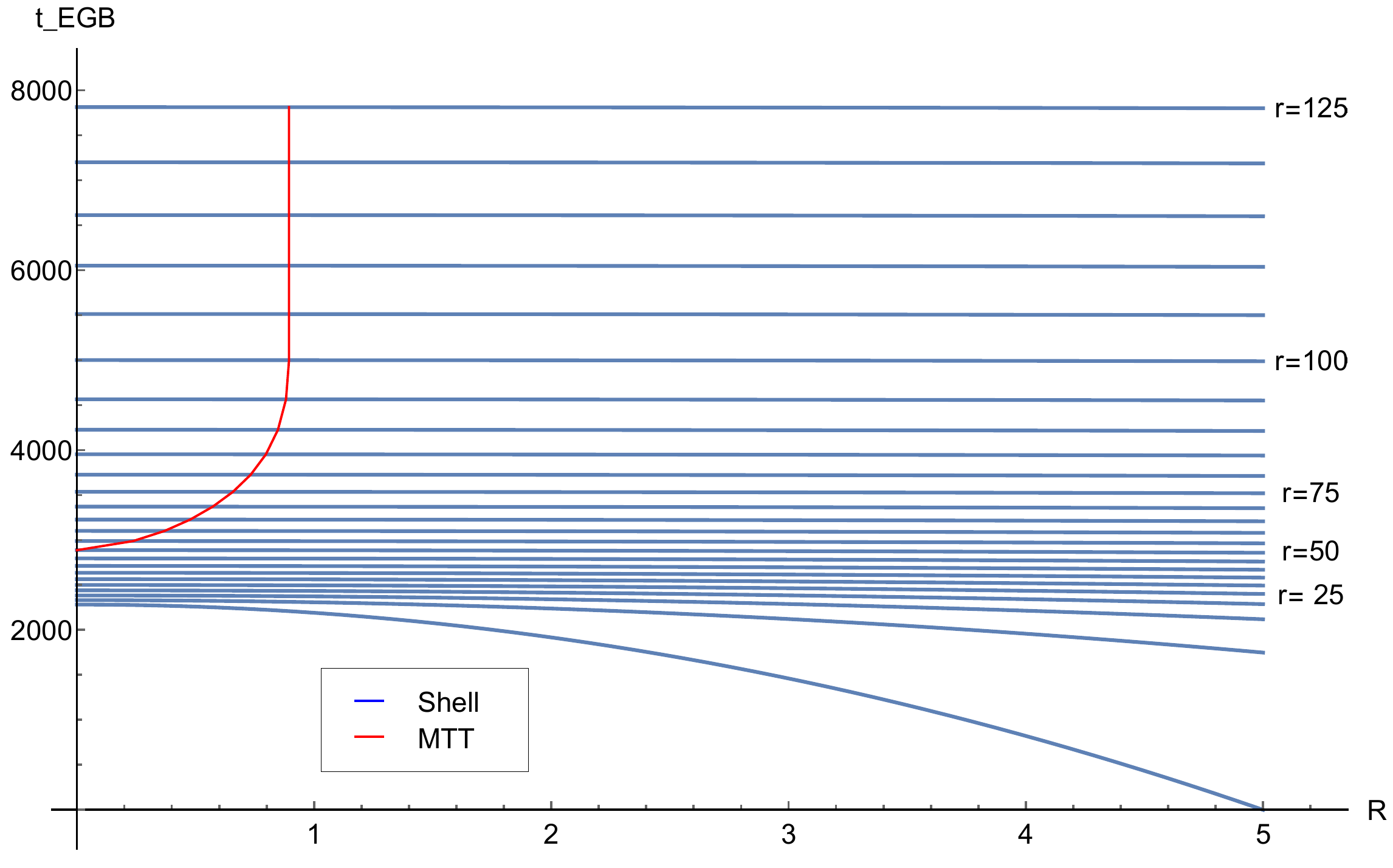}}
\caption{\label{fig:ltbk0_example2EGB}The graphs show the (a) density distribution $\rho$, for equation 
\eqref{rhoLTBEGB}, (b) values of $C$,
and (c) formation of MTT along with the shells. For the plot we have used $\lambda=0.1$.
Note that the MTT begins to form only after some shells have fallen in the singularity.
This is a direct consequence of the fact that eqn. \eqref{trapped_surface_eqn} requires
the mass function $F(r)$ to exceed $2\lambda$ for a real valued $R_{M}(r,t)$ in the equation of MTS.}
\end{figure}
%

%
\item Let us now consider a Gaussian density profile with 
the density given by the following form:
\begin{equation}\label{rho_gaussian}
\rho(r)=\frac{3m_{0}}{r_{0}^{4}}\exp (-r^{2}/r_{0}^{2}),
\end{equation}
where $m_{0}$ is the total mass of the matter cloud, $r_{0}$ is a parameter which indicates 
the distance where the density of the cloud decreases to $[\rho\,(0)/e]$. 
%
\begin{figure}[t!]
\centering
\subfigure[\ ] {
\includegraphics[scale=0.4]{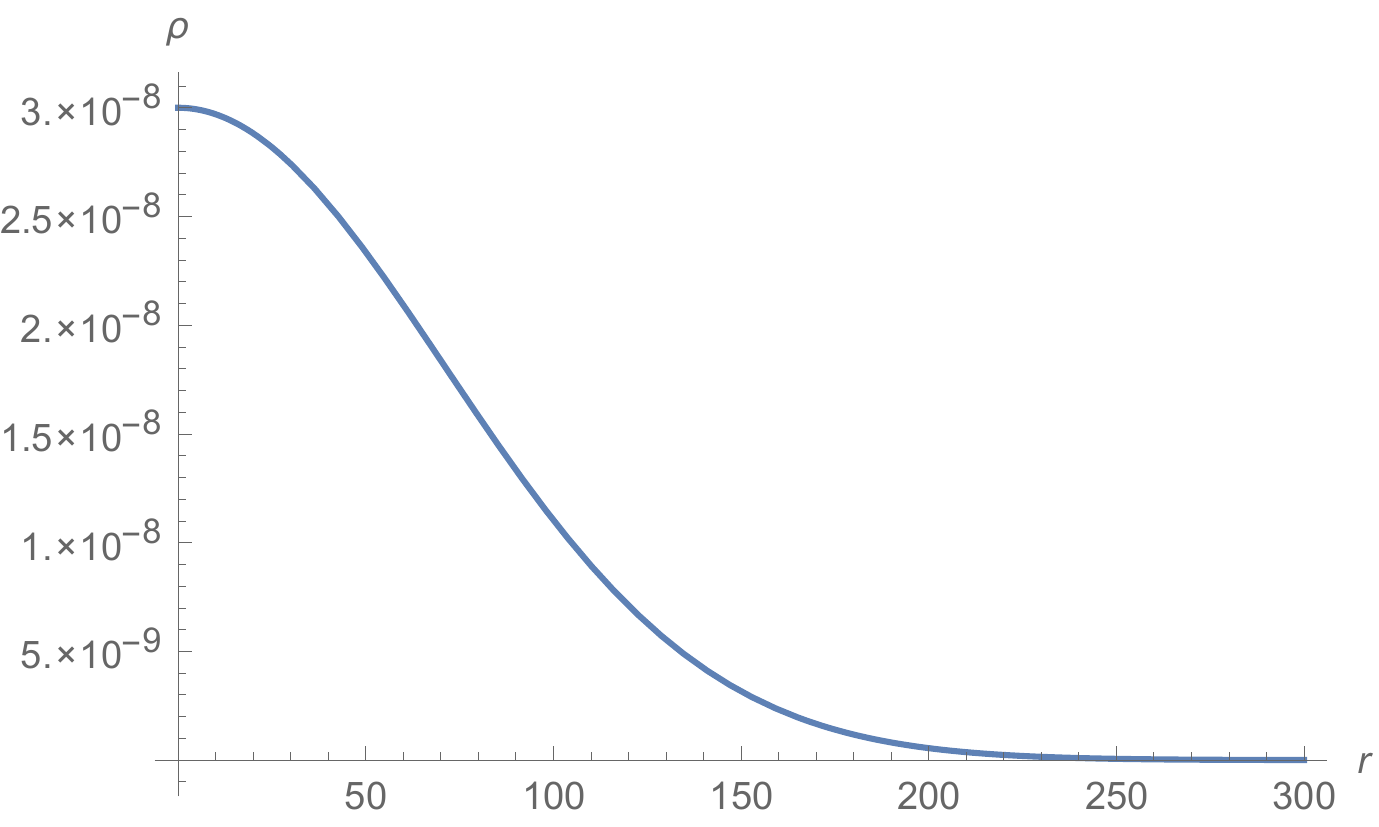}}
\subfigure[\ ] {
\includegraphics[scale=0.4]{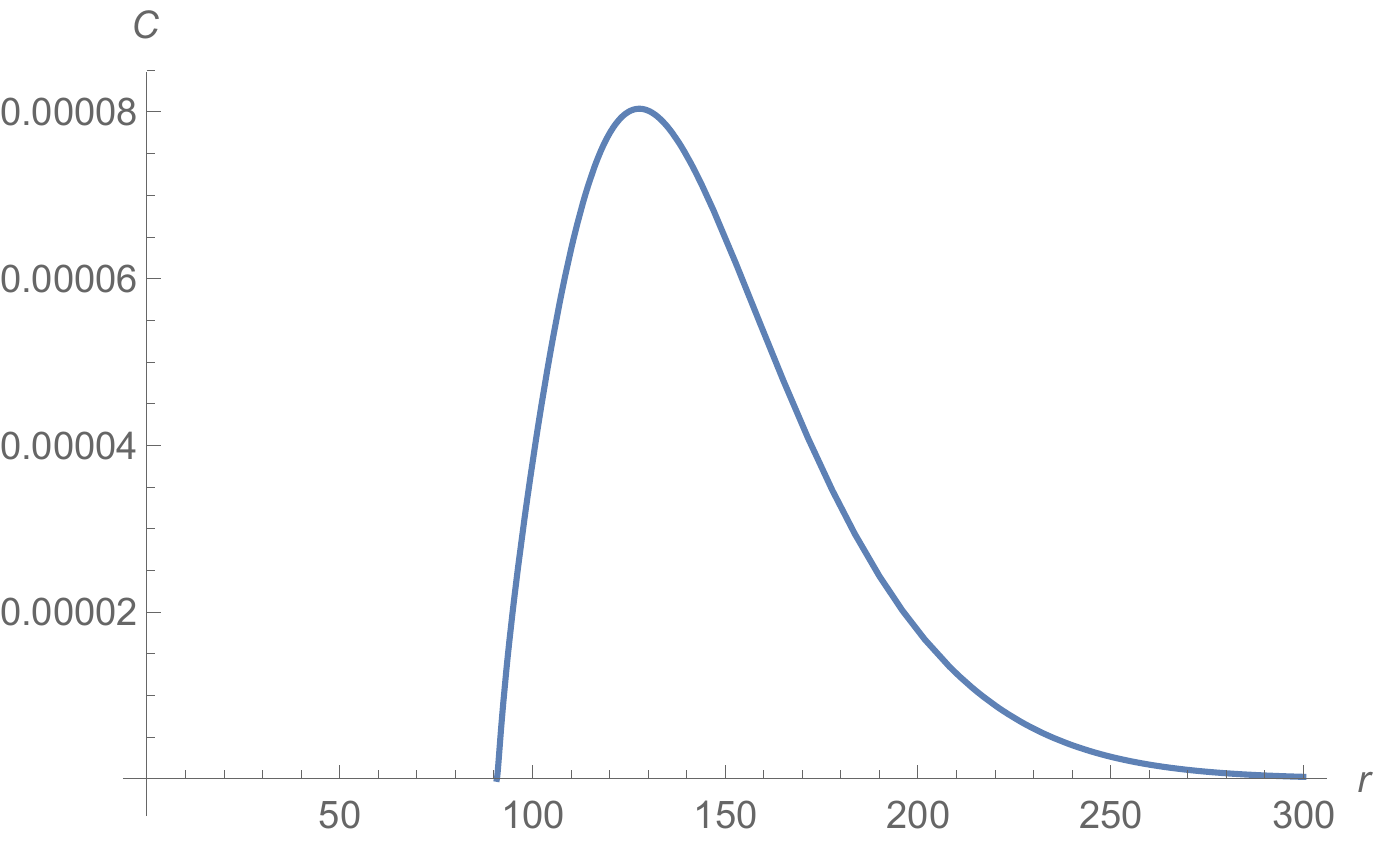}}
\subfigure[\ ] {
\includegraphics[scale=0.5]{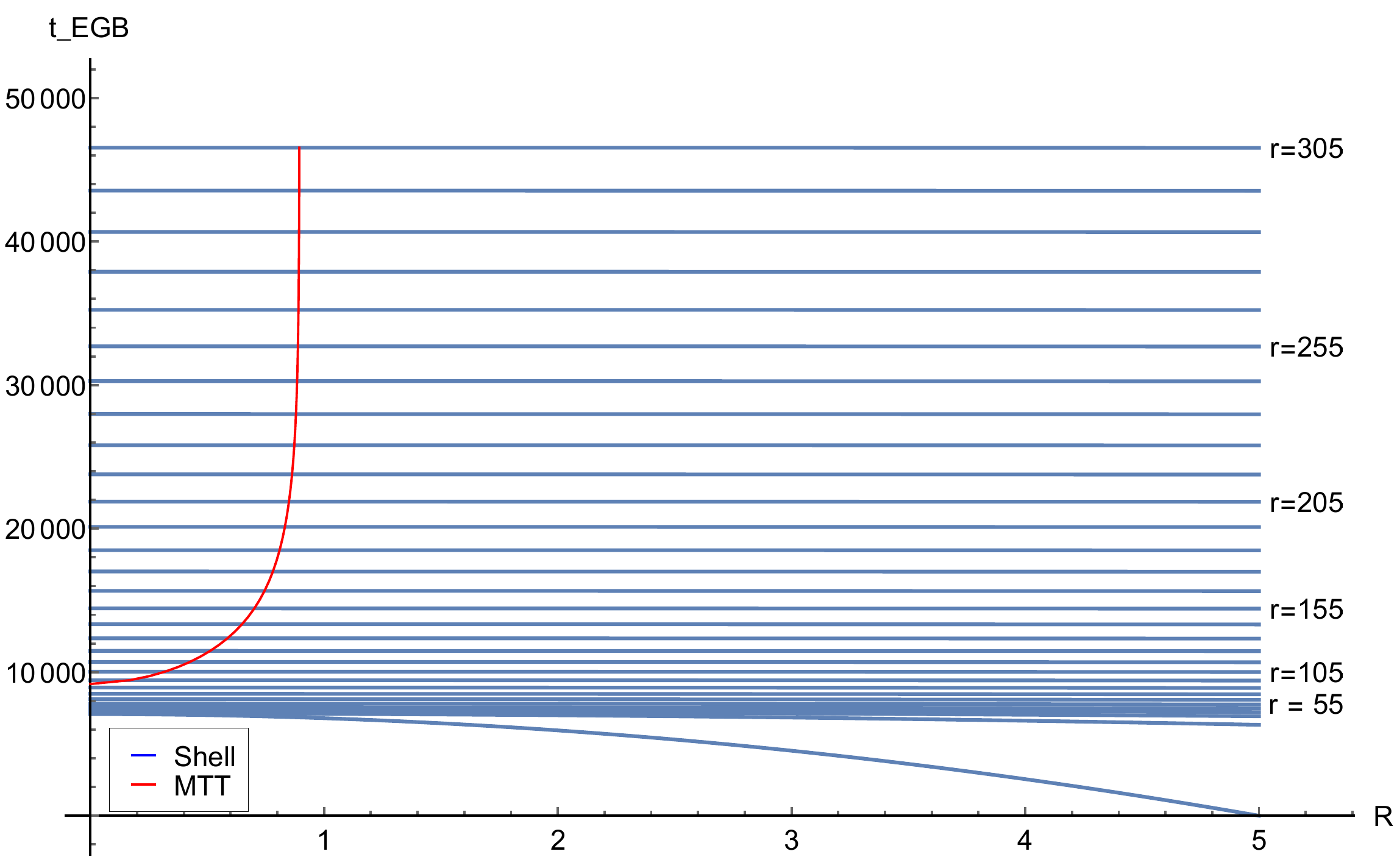}}
\caption{\label{fig:Gaussk0_example2EGB}
The graphs show the (a) density distribution for eqn. \eqref{rho_gaussian}, (b) values of $C$,
and (c) formation of MTT along with the shells. 
The straight lines of MTT in (c), after the shell $r=300$, represents the isolated horizon phase.}
\end{figure}
%
In our example,
we have chosen $r_{0}=100\, m_{0}$ and the EGB coupling constant $\lambda=0.1$.
Note that the MTT begins only after the shell at $r=90$ has fallen in. As
explained in the previous subsection, this is a direct
consequence of the relation eqn.\eqref{trapped_surface_eqn}. 
The MTT in Fig. \eqref{fig:Gaussk0_example2EGB}(b) and \eqref{fig:Gaussk0_example2EGB}(c)
clearly shows that the MTT is spacelike, and attains the IH phase when the shells
at $r=300$ has fallen in.

%
\item Let us consider a density profile given by the following form for 
$r\in[0, \pi r_{0}]$:
\begin{equation}\label{timelike_rho}
\rho(r)=(\gamma/r_{0}^{2})\left[\pi-(r/5r_{0})\{3+2\cos^{2}(5r/r_{0})\}\right]
\end{equation}
where $\gamma$ is a dimensionless constant.  This example constitutes a situation where
the MTTs are a series of timelike membranes interspaced with dynamical horizons.
A similar profile was used to study gravitational collapse in $4$d 
GR \cite{Booth:2005ng, Chatterjee:2020khj}.
%
\begin{figure}[h!]
\centering
\subfigure[\ ] {
\includegraphics[scale=0.4]{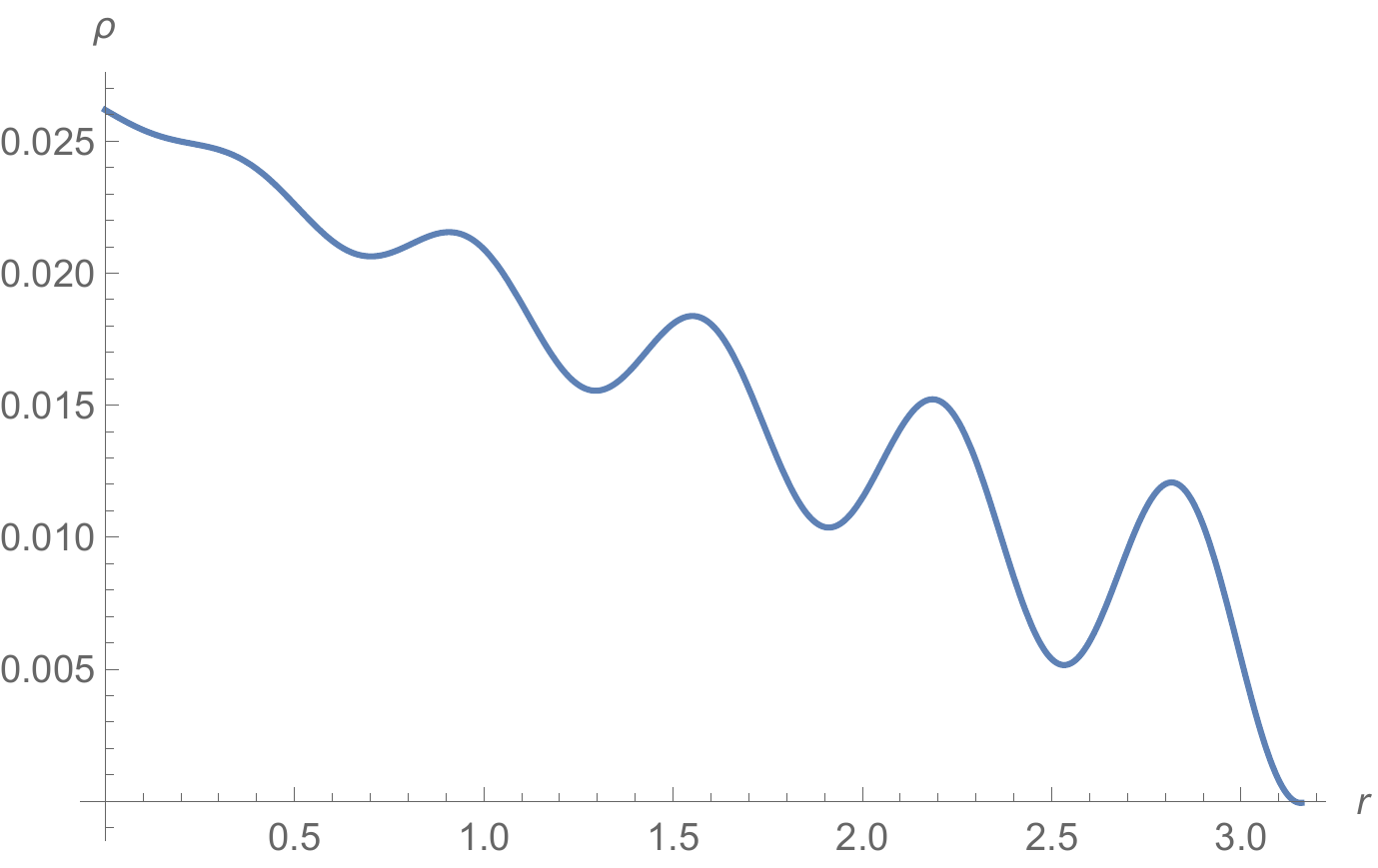}}
\subfigure[\ ] {
\includegraphics[scale=0.4]{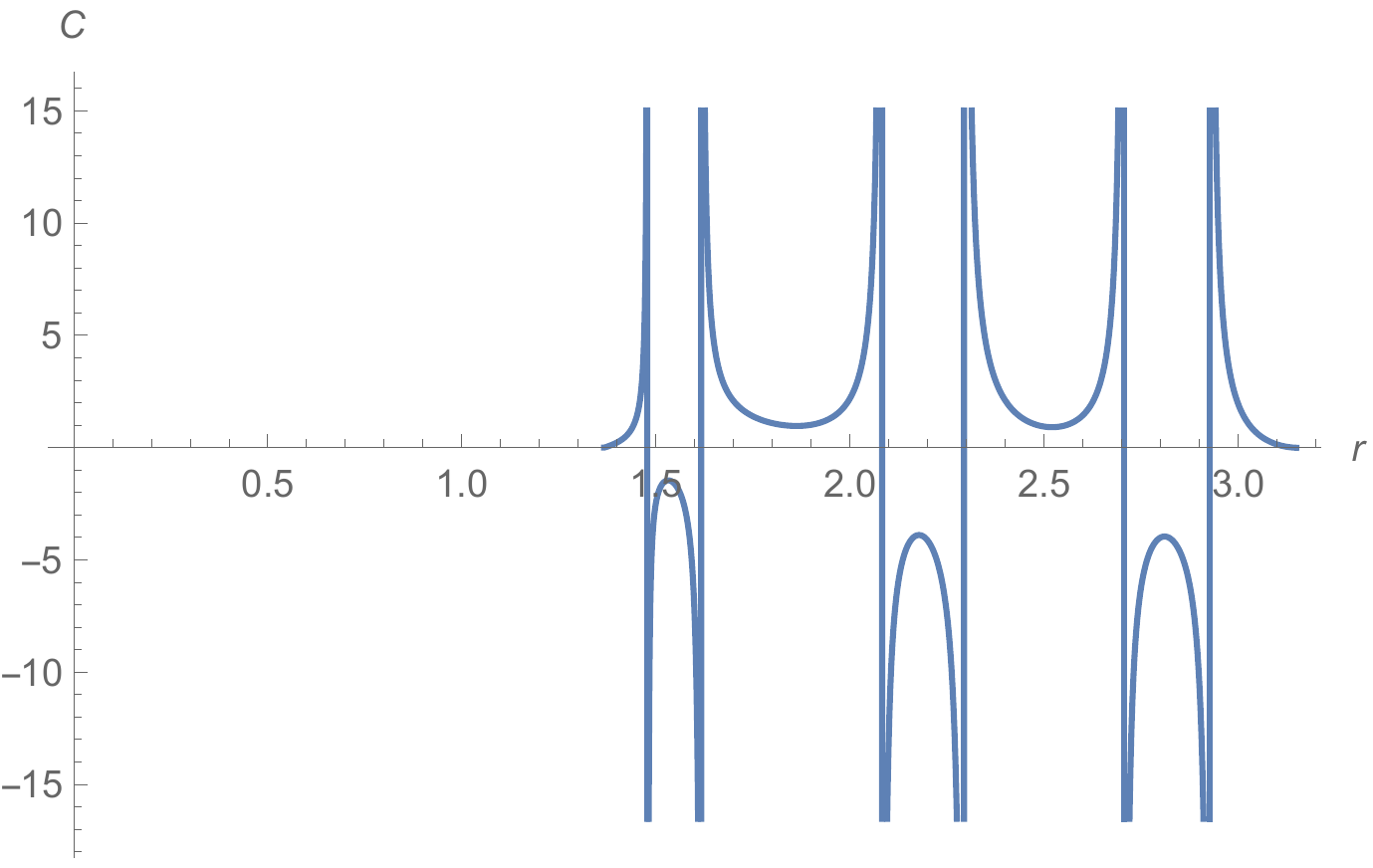}}
\subfigure[\ ] {
\includegraphics[scale=0.4]{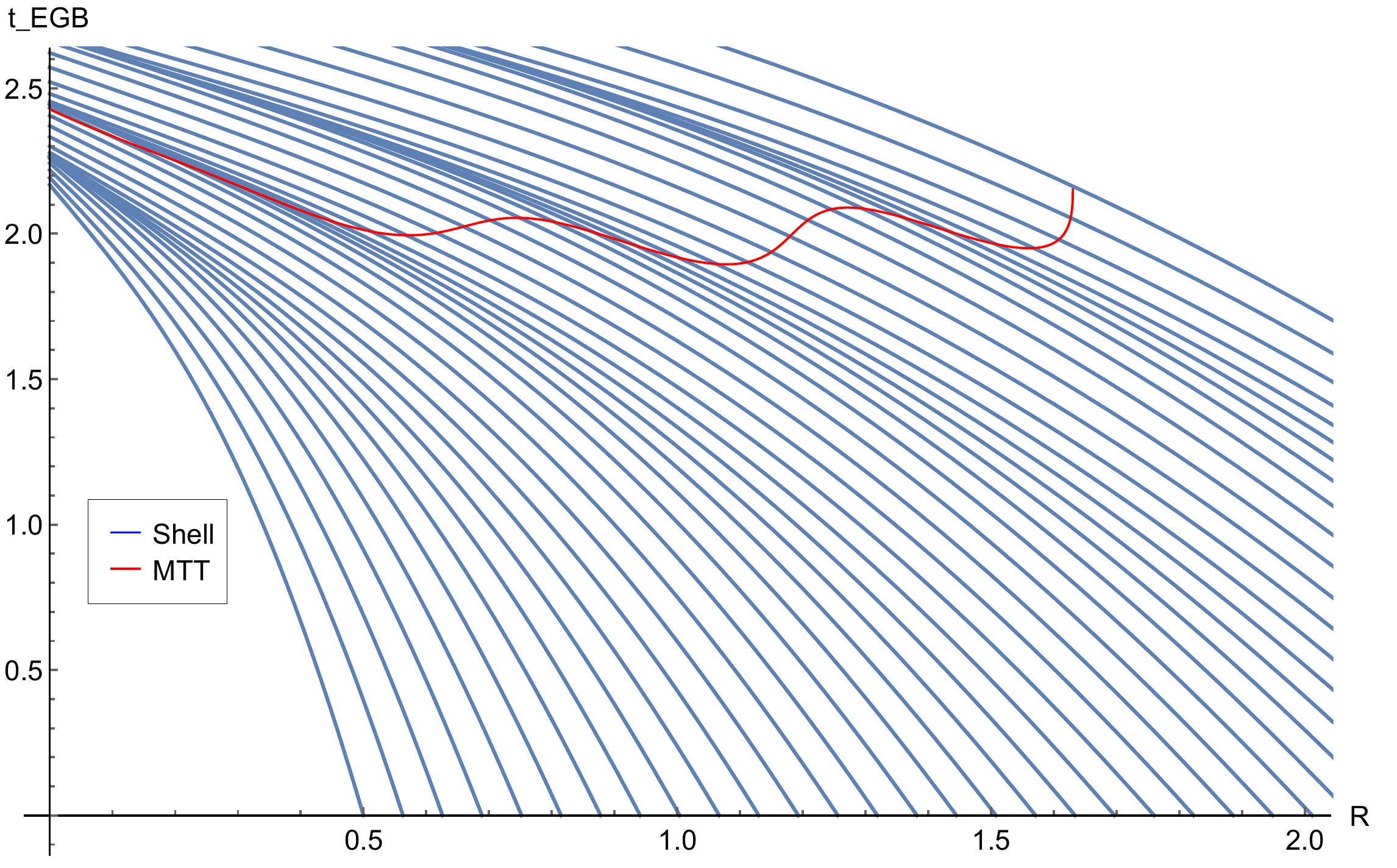}}
\caption{\label{fig:Timelikek0_example2EGB} The graphs show the (a) density distribution, (b) values of $C$,
and (c) formation of MTT along with the shells for the density in equation \eqref{timelike_rho}.}
\end{figure}
%
In our example,
we have chosen $r_{0}=1$ , $\gamma=1/120$, and the EGB coupling constant $\lambda=0.1$.
Note the peculiar dynamics of the MTT  from
figure \eqref{fig:Timelikek0_example2EGB}(c). The MTT first forms for the shell at $r=1.5$,
and then evolves in a timelike manner to reach towards the MTT formed after the shells at $r=1.35$
have fallen in. Again note that during the initial period, the central singularity
is not covered by the MTT and remains naked, as expected due to equation \eqref{trapped_surface_eqn}. 
During the period the shells from $r=1.7$ to $r=2.0$ collapse, the MTT is a dynamical horizon,
as may also be confirmed from the graph of $C$ in figure \eqref{fig:Timelikek0_example2EGB}(b).
This behaviour is repeated until matter stops falling at $r=3.0$, when the MTT reaches 
the equilibrium state of an IH.

%
\item Two shells falling consecutively on a black hole:
Let us assume that a black hole of mass $M$ exists, upon which a density profile of the following form
falls:
\begin{equation}\label{rho_MAX}
\rho(r)=\frac{12\,(m_{0}/\,r_{0}^{4}\,)\,[(r/r_{0})-\varsigma]^{\,2}}{[2\left( 4+\varsigma^{2}\right) +(9+2\varsigma^{2})\sqrt{\pi}e^{\varsigma^{2}}\{1+\varsigma\erf(\varsigma)\}]}
\, \exp[(2r/r_{0})\varsigma-(r/r_{0})^{\,2}\,],
\end{equation}
where $m_{0}=M/2$, ($M=1)$ is the mass of the shell, $2r_{0}$ is the width of each shell,
and $\varsigma=10m_{0}$. 
%
\begin{figure}[h!]
\centering
\subfigure[\ ] {
\includegraphics[scale=0.4]{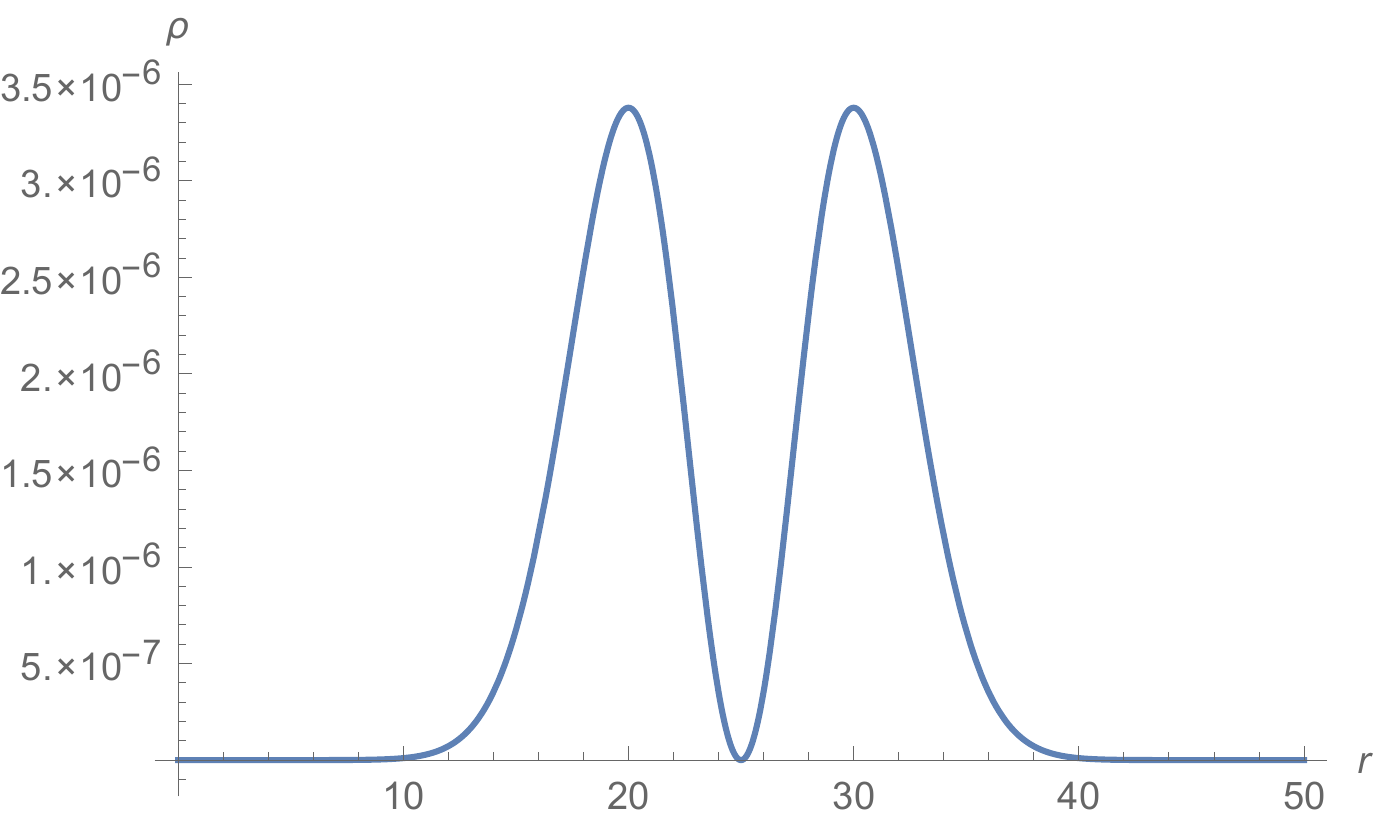}}
\subfigure[\ ] {
\includegraphics[scale=0.4]{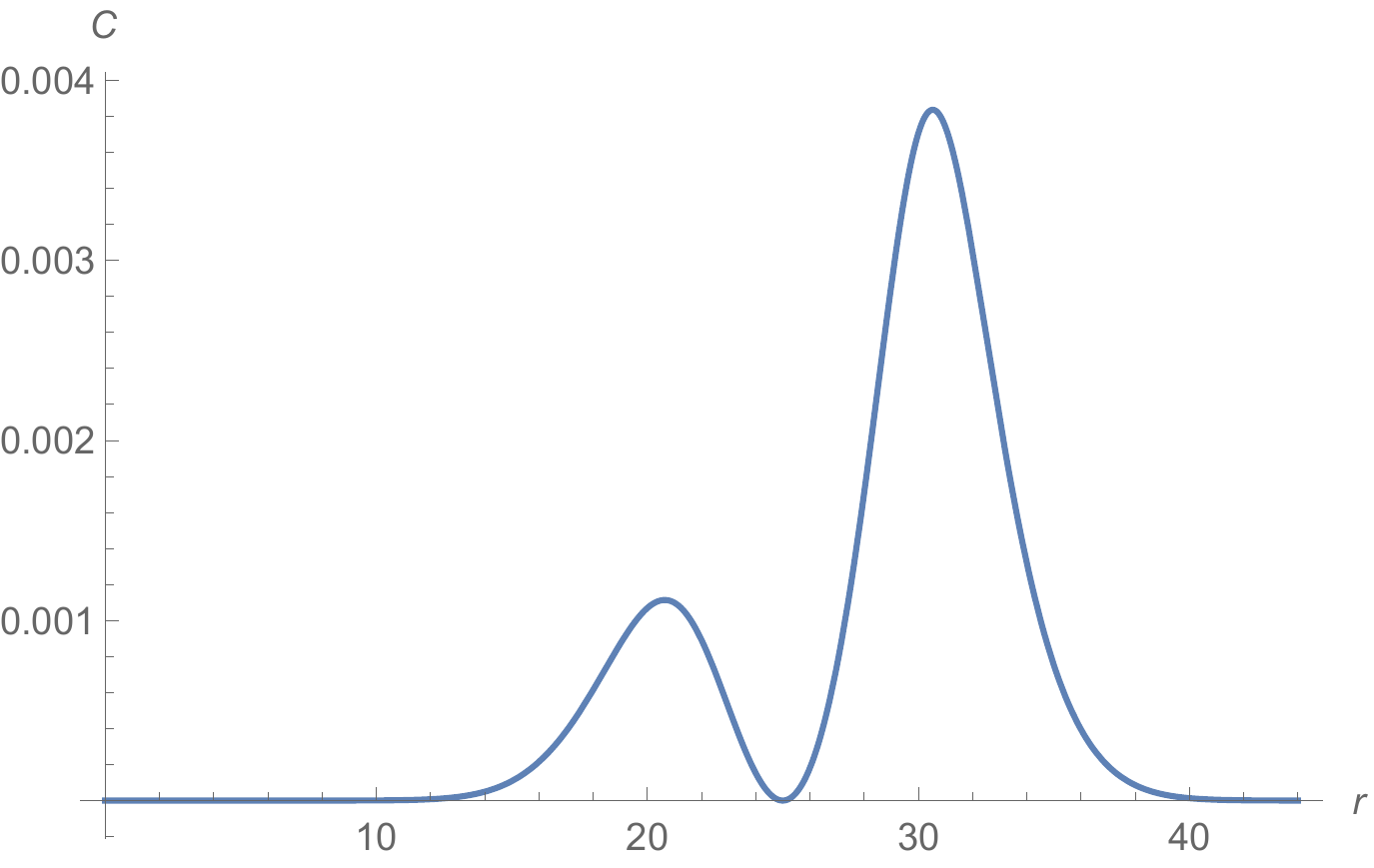}}
\subfigure[\ ] {
\includegraphics[scale=0.4]{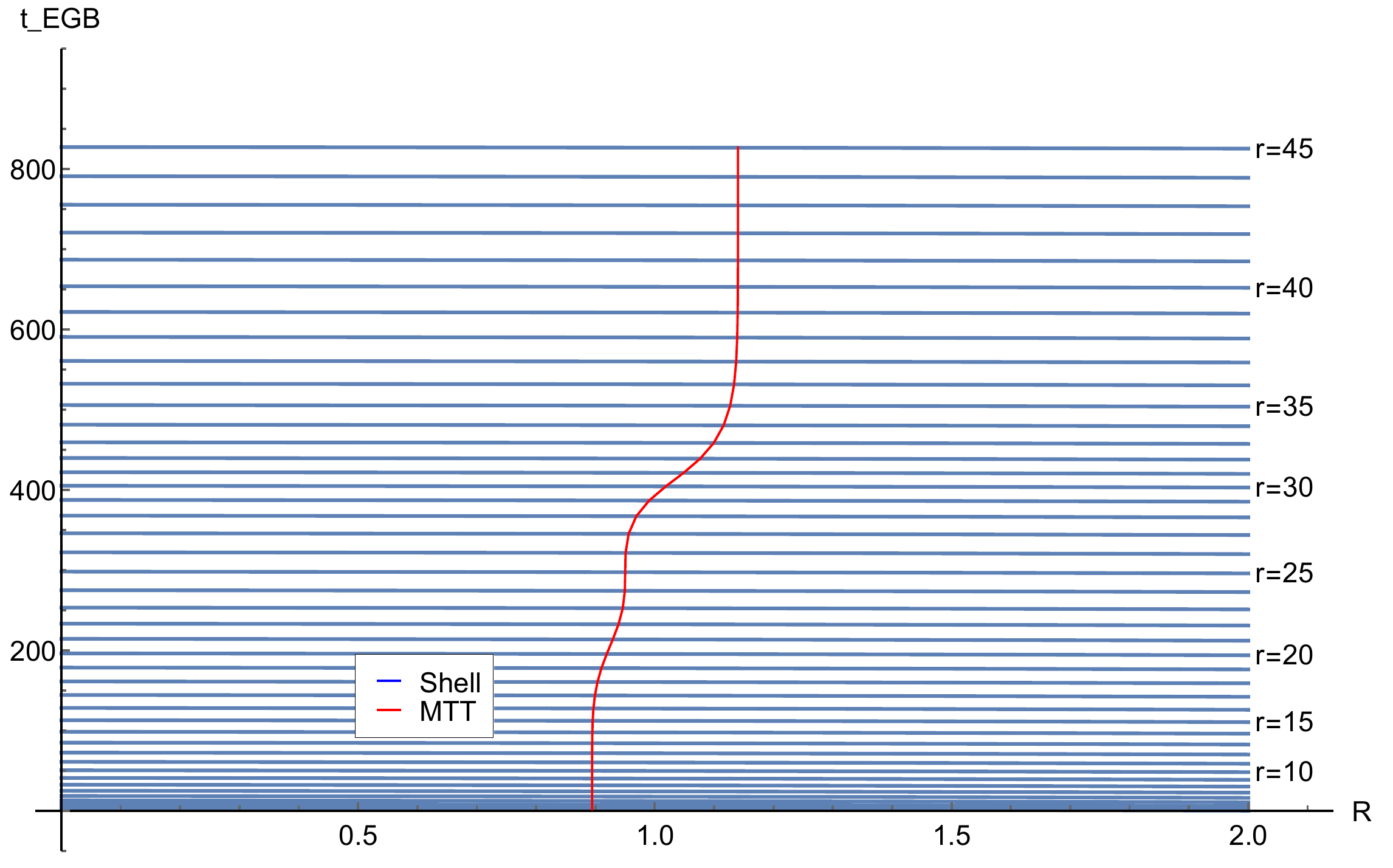}}
\caption{\label{fig:Maxk0_example4EGB}
The graphs show the (a) the density profile from eqn.\eqref{rho_MAX}, (b) values of $C$,
and (b) formation of MTT along with the shells which fall consecutively 
on a black hole.}
\end{figure}
%
The graphs corresponding to this case is given in \eqref{fig:Maxk0_example4EGB}.
Note that these graphs constitute the case where two mass profiles fall on a black hole
one after the other. The spacetime singularity already exists into which these shells
fall in. Note that as the fist profile falls, the MTT begins from the already existing horizon 
at $R=0.89$ and develops until the shells corresponding to $r=23$ to $r=25$ fall in carrying no
mass with them. At these times, the MTT reaches an equilibrium state, and becomes dynamical only
after the second mass profile begins to fall. So, the MTT passes through multiple stages
of dynamical horizon, interspaced with isolated horizons when no matter is infalling.
This behaviour is easily verifiable 
from figures \eqref{fig:Maxk0_example4EGB}(b), and \eqref{fig:Maxk0_example4EGB}(c).


 \end{enumerate}
%

%
%

\subsection{Bounded collapse}
For bounded collapse, we again have $\alpha'=0$ and  $G(r,t)=e^{-2\beta(r, t)}R'^2=1-k(r)=E(r)$, 
where $E(r)$ is the integration function.
For this case $k>0$, the equation of motion is given by \eqref{EQMEGB}
\begin{eqnarray}
\dot{R}\,^{2}(r,t)&=&-k(r)-\frac{R^2(r,t)}{4\lambda}+\frac{R^2(r,t)}{4\lambda}\left[1+\frac{8\lambda F(r,t)}{R^4(r,t)}\right]^{1/2}. \label{dotRkg0EGB}
\end{eqnarray}
This equation of motion (\ref{dotRkg0EGB}) can be rewritten in the following form: 
\begin{eqnarray}
dt=-\frac{2\sqrt{\lambda}\, dR}{\sqrt{-R^2-4\lambda k+\sqrt{R^4+8\lambda F}}} \label{EOMk}
\end{eqnarray}
To integrate this equation of motion \eqref{EOMk}, we consider a parametric choice 
of $R(r,t)$ of the following form:
\begin{equation}
 x=-R^{\,2}(r,t)-4\lambda k(r)+\sqrt{R^{\,4}(r,t)+8\lambda F(r)}
\end{equation}
A simple calculation of squaring both sides leads to the following expression: 
\begin{eqnarray}\label{Rparametric}
R(r,t)=\frac{1}{\sqrt{2}}\left[\frac{8\lambda F}{(x+4\lambda k)}-(x+4\lambda k)\right]^{1/2}
\end{eqnarray}
Using this expression of eqn. \eqref{Rparametric}, a simple calculation leads to
modification of \eqref{EOMk}:
\begin{eqnarray}
dt=\frac{\sqrt{\lambda}\left( (x+4\lambda k)^2+8\lambda F\right) dx}{\sqrt{2}\sqrt{x}(x+4\lambda k)^{3/2}\sqrt{8\lambda F-(x+4\lambda k)^2}}
\end{eqnarray}
The integration of the above equation gives the equation of the collapsing shell to be:
\begin{eqnarray}
t_{sh}&=&A_{1}\left[\left(8\lambda F-2\sqrt{2\lambda F}(x+4\lambda k)\right)+A_{2}\left\{\left(\sqrt{2F}+2\sqrt{\lambda}k\right)\, \text{EllipticE}[N_{1},2 N_{2}]\right.\right.\nonumber\\
&&\left.\left.-\left(\sqrt{2F}-2\sqrt{\lambda}k\right)\, \text{EllipticF}[N_{1},2 N_{2}]-2\sqrt{\lambda} k\, \text{EllipticPi}[N_{2},N_{1},2 N_{2}]\right\}\right]\nonumber\\
&-&(A_{1})_{0}\left[\left(8\lambda F-2\sqrt{2\lambda F}(x_{0}+4\lambda k)\right)
+(A_{2})_{0}\left\{\left(\sqrt{2 F}+2\sqrt{\lambda}k\right)\, \text{EllipticE}[(N_{1})_{0},2 N_{2}]\right.\right.\nonumber\\
&&\left.\left.-\left(\sqrt{2F}-2\sqrt{\lambda}k\right)\, \text{EllipticF}[(N_{1})_{0},2 N_{2}]-2\sqrt{\lambda} k\, \text{EllipticPi}[N_{2},(N_{1})_{0},2 N_{2}]\right\}\right]\label{tshell}
\end{eqnarray}
The equation for the spherical MTTs, are obtained for $R(r,t)=\sqrt{F(r)-2\lambda}$ and gives:
\begin{eqnarray}
t_{AH}&=& (A_{1})_{2M}\left[\left(8\lambda F-2\sqrt{2\lambda F}(x_{2m}+4\lambda k)\right)
+(A_{2})_{2M}\left\{\left(\sqrt{2F}+2\sqrt{\lambda}k\right)\, \text{EllipticE}[(N_{1})_{2M},2 N_{2}]\right.\right.\nonumber\\
&&\left.\left.-\left(\sqrt{2F}-2\sqrt{\lambda}k\right)\, \text{EllipticF}[(N_{1})_{2M},2 N_{2}]-2\sqrt{\lambda} k\, \text{EllipticPi}[N_{2},(N_{1})_{2M},2 N_{2}]\right\}\right]\nonumber\\
&&-(A_{1})_{0}\left[\left(8\lambda F-2\sqrt{2\lambda F}(x_{0}+4\lambda k)\right)
+(A_{2})_{0}\left\{\left(\sqrt{2F}+2\sqrt{\lambda}k\right)\, \text{EllipticE}[(N_{1})_{0},2 N_{2}]\right.\right.\nonumber\\
&&\left.\left.-\left(\sqrt{2F}-2\sqrt{\lambda}k\right)\, \text{EllipticF}[(N_{1})_{0},2 N_{2}]-2\sqrt{\lambda} k\, \text{EllipticPi}[N_{2},(N_{1})_{0},2 N_{2}]\right\}\right],\label{tapp}
\end{eqnarray}
where the coefficients $A_{1}, \, A_{2}$, and the arguments $N_{1},\, N_{2}$ are given by:
\begin{eqnarray*}
	A_{1}=\frac{\sqrt{x}}{k(x+4\lambda k)^{3/2}\sqrt{8\lambda F-(x+4\lambda k)^2}}, \,\,&&
	A_{2}=\frac{(2^{5}\lambda^{2}F)^{1/4} \sqrt{x+4\lambda k} \sqrt{-8\lambda F+(x+4\lambda k)^2}}{\sqrt{x\sqrt{2F}+2\sqrt{\lambda}k}}\\
N_{1}=\sin^{-1}\left[\frac{(\sqrt{2F}+2\sqrt{\lambda}k)(x+4\lambda k)}{2\lambda k \{2\sqrt{2\lambda F}+(x+4\lambda k)\}}\right]^{1/2}, &&
N_{2}=\frac{2\sqrt{\lambda}k}{\sqrt{2F}+2\sqrt{\lambda}k}.\\
	\end{eqnarray*}
The terms with subscript $0$ and $2M$ represents its value at 
the initial shells at $r=r_{0}$ and at the formation of MTT with 
$R=R=\sqrt{F-2\lambda}$. For example, $x=-R^2-4\lambda k+\sqrt{R^4+8\lambda F}$,
whereas, its value at $0$ represents $x_{0}=-r^2-4\lambda k+\sqrt{r^4+8\lambda F}$. 
The EllipticF[$\phi,\, z$] represents an incomplete elliptic integral
of the first kind, EllipticE[$\phi,\, z$] is elliptic integral of the second kind,
whereas EllipticPi[$y;\,\phi,\, z$] is an elliptic integral of the third 
kind \cite{Whittaker_Watson}.

 In the following we shall take several examples to show how a MTT develops during the bounded gravitational collapse in EGB theory.
%
%
\subsubsection*{Example}
%
\begin{enumerate}

\item Let us consider a Gaussian profile with the density given by eqn. \eqref{rho_gaussian}. 
The form of the density is same as in  figure \eqref{fig:Gaussk0_example2EGB}(a).
%
\begin{figure}[t!]
\centering
\subfigure[\ ] {
\includegraphics[scale=0.4]{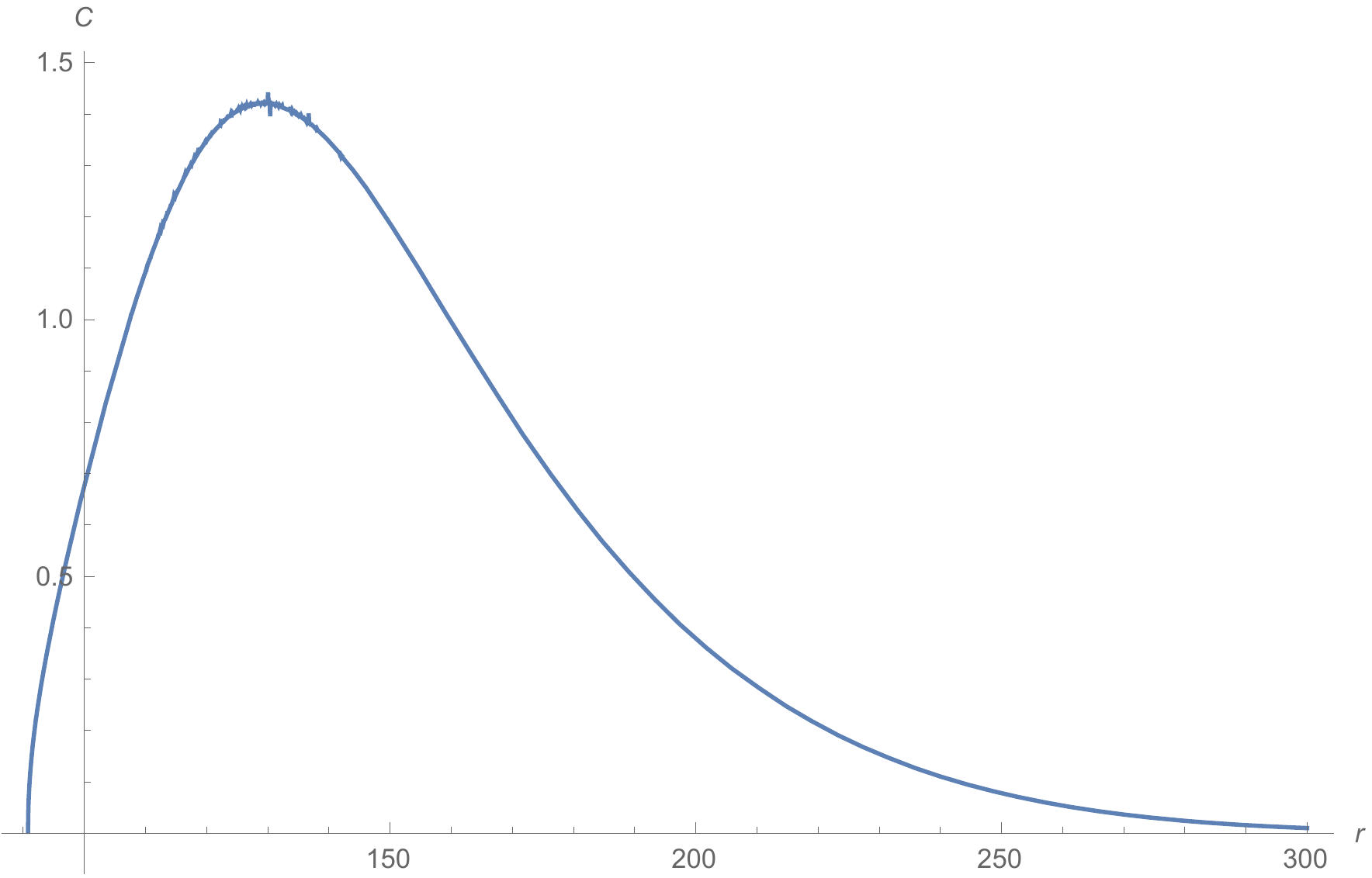}}
\subfigure[\ ] {
\includegraphics[scale=0.4]{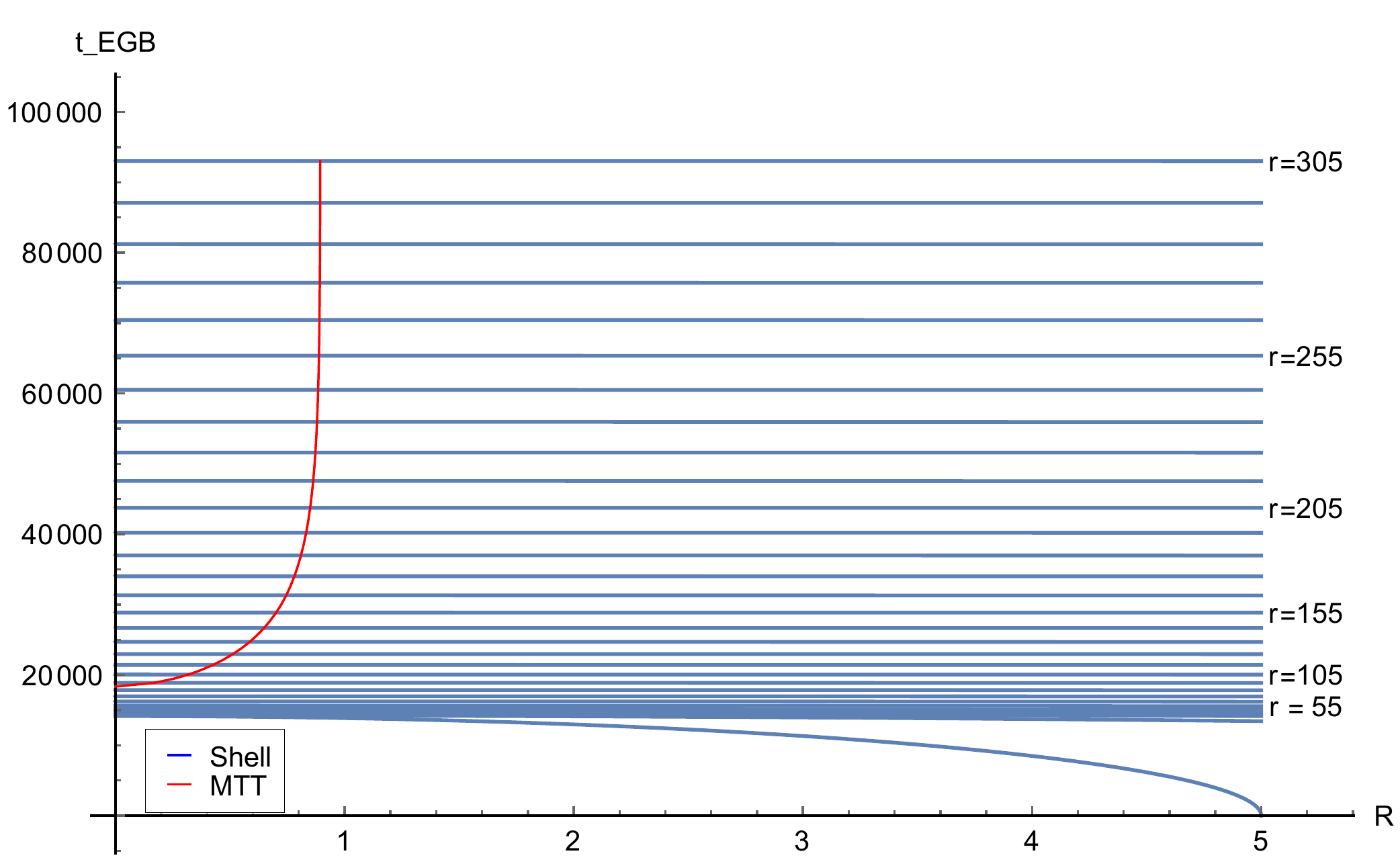}}
\caption{\label{fig:Gausskg0_example2EGB} The graphs show the (a) values of $C$,
and (b) formation of MTT along with the shells
for the Gaussian distribution of eqn. \eqref{rho_gaussian}.}
\end{figure}
%
In our example,
we have chosen $r_{0}=100\, m_{0}$. The behaviour of MTT is similar to 
that discussed for the marginally bound case, see figure \eqref{fig:Gausskg0_example2EGB}.
However, the time of formation of MTT and the value of the $C$ has changed in comparison. 
Again note that the MTT forms only after shells at $r=90$ collapse.
Before that shell falls in, the singularity remains naked. The time 
of formation of MTT changes in comparison to 
the marginally bound case of figure \eqref{fig:Gaussk0_example2EGB}.
The straight lines of MTT in (b), after the shell $r=300$, represents the isolated horizon phase.


\item Let us consider a density profile given by the following form: 
\begin{equation}\label{expkg0}
\rho(r)=\left(m_{0}/8\pi r_{0}^{4}\right)\exp (-r /r_{0}),
\end{equation}
where $m_{0}$ is the total mass of the matter cloud, $r_{0}$ is a parameter which indicates 
the distance where the density of the cloud decreases to $[\rho\,(0)/e]$. The MTT begins
after the shells at $r=30$ have fallen into the singularity. The MTT remains spacelike
through out its time evolution, and reaches an equilibrium state only after the 
the density reaches negligible values. These conclusions are easily be verified from the 
graphs in figure \eqref{fig:expkg0_example2EGB}. Note again that
the MTT begins only after sufficient number of shells have collapsed to the singularity
in accordance to the choice of $\lambda=0.1$ in eqn. \eqref{trapped_surface_eqn}. 
%
\begin{figure}[t!]
\centering
\subfigure[\ ] {
\includegraphics[scale=0.35]{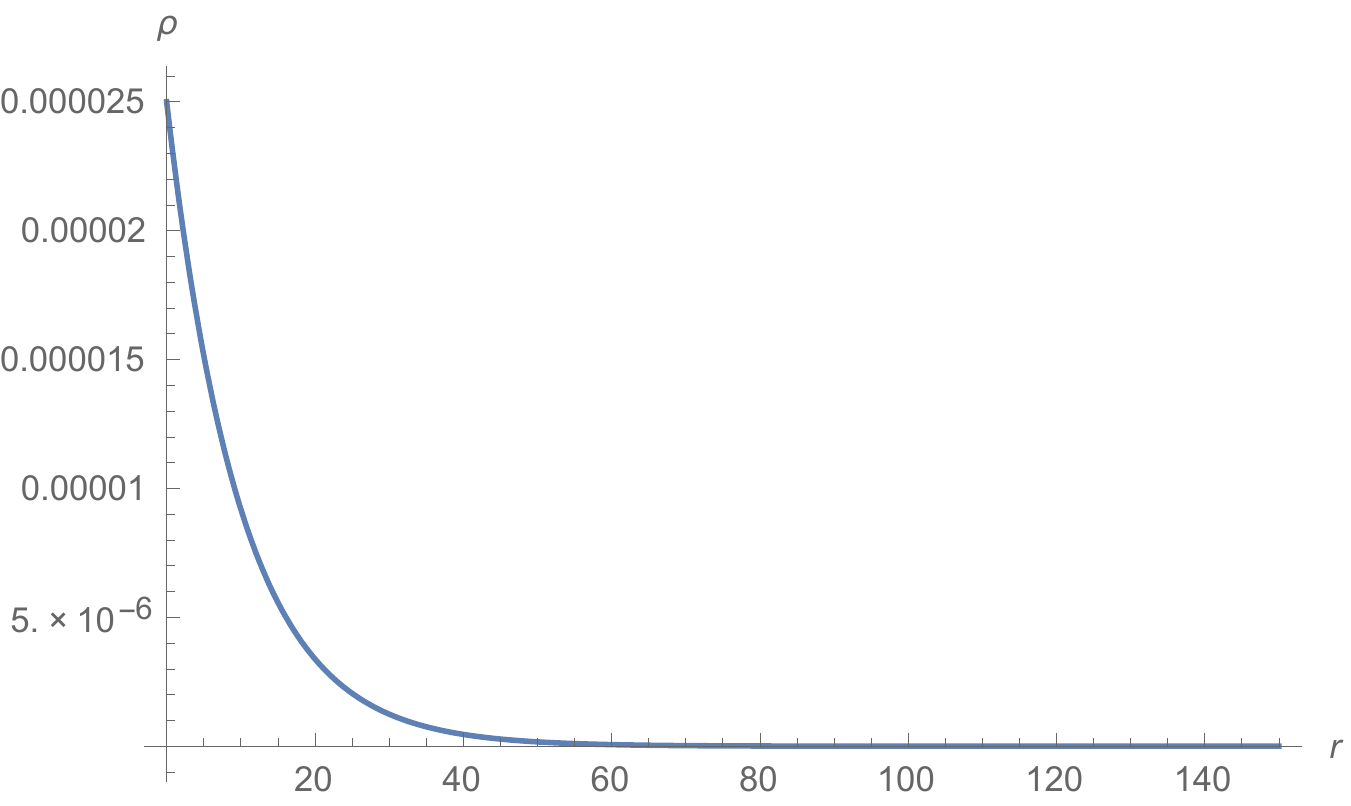}}
\subfigure[\ ] {
\includegraphics[scale=0.35]{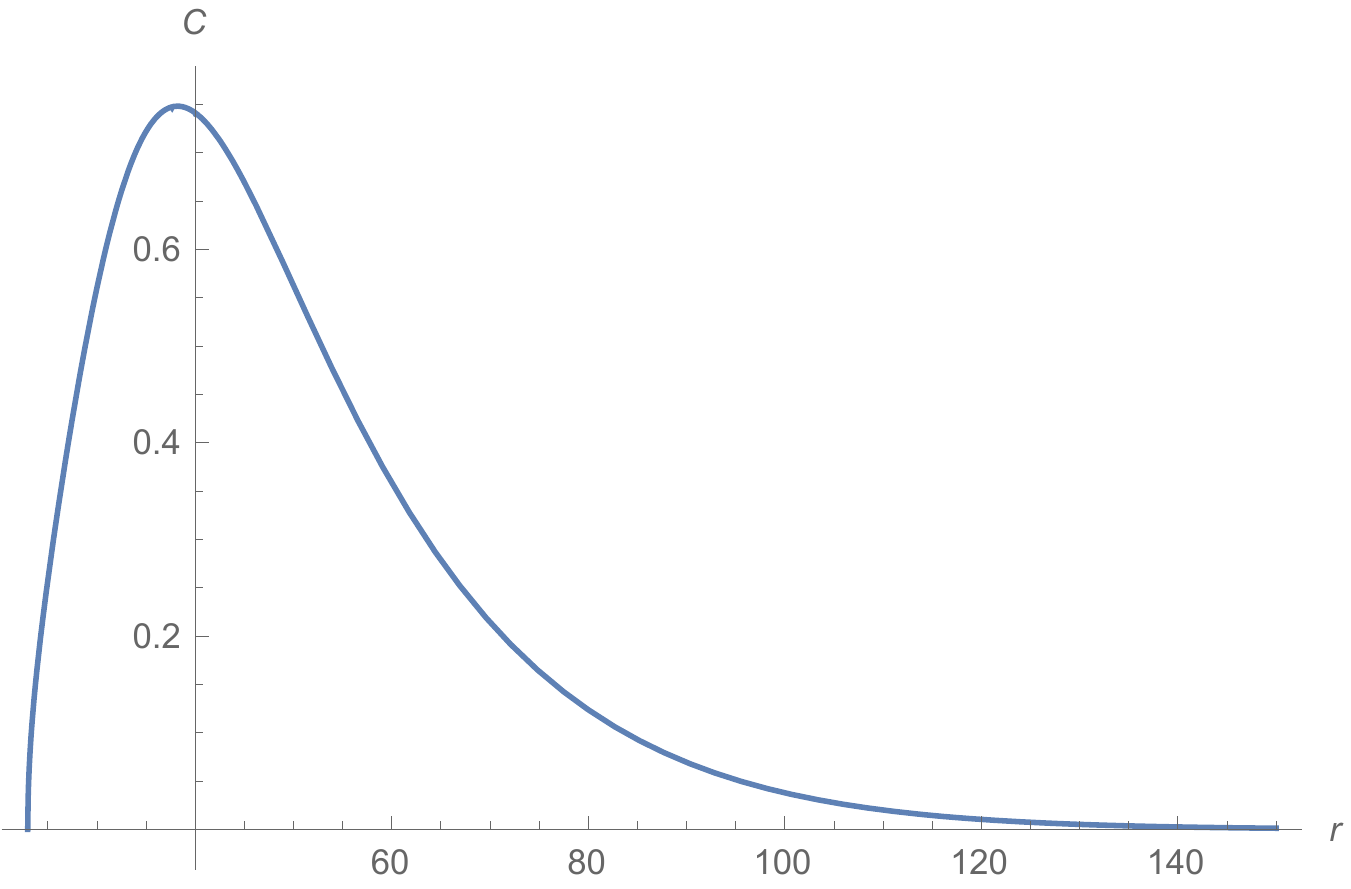}}
\subfigure[\ ] {
\includegraphics[scale=0.3]{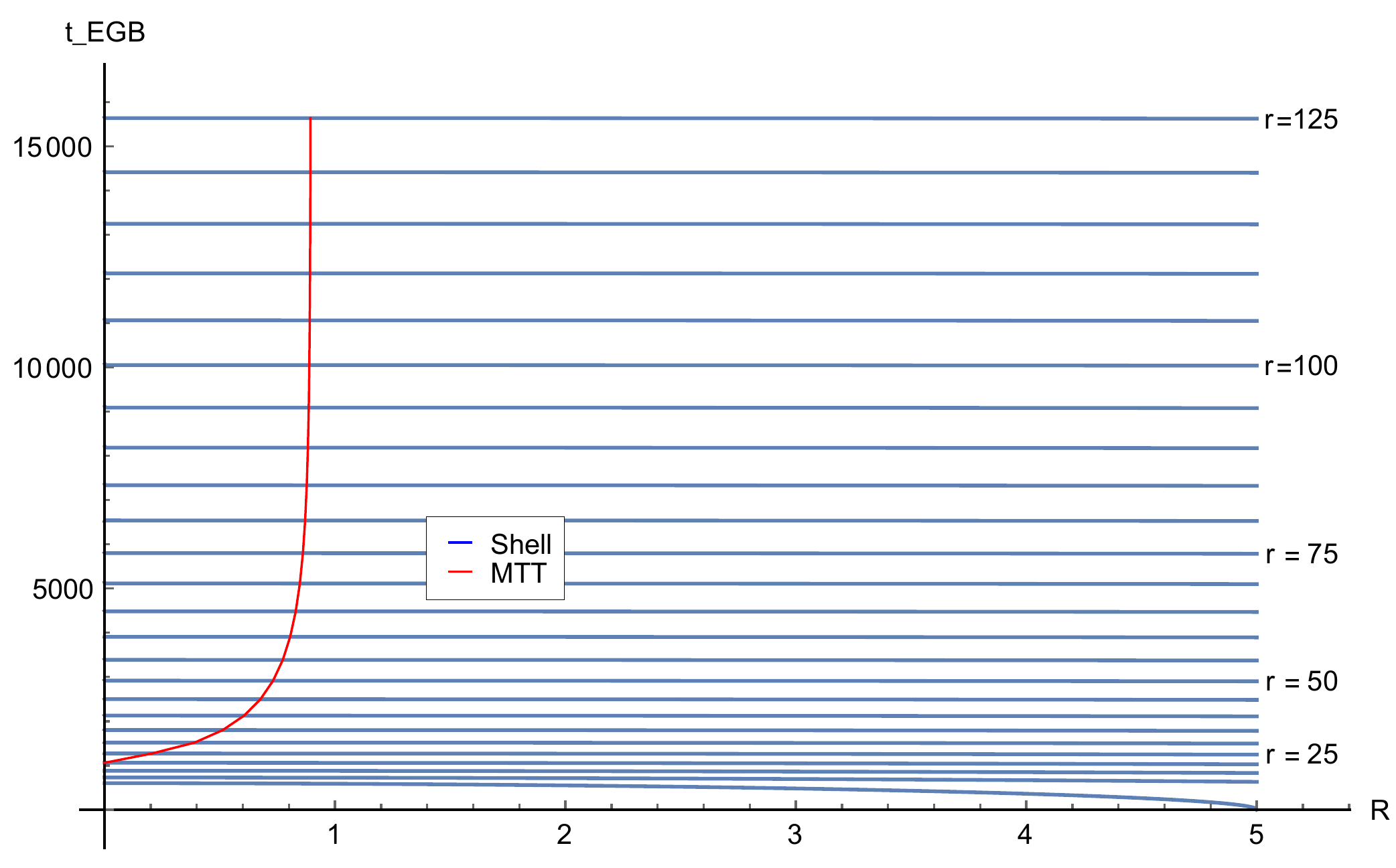}}
\caption{\label{fig:expkg0_example2EGB}The graphs show the (a) values of $C$,
and (b) formation of MTT along with the shells
for the matter profile with exponentially falling density distribution given in eqn. \eqref{expkg0}. 
The MTT is spacelike.}
\end{figure}
%


\item Two shells falling consecutively on a black hole:
%
%
\begin{figure}[t!]
\centering
\subfigure[\ ] {
\includegraphics[scale=0.4]{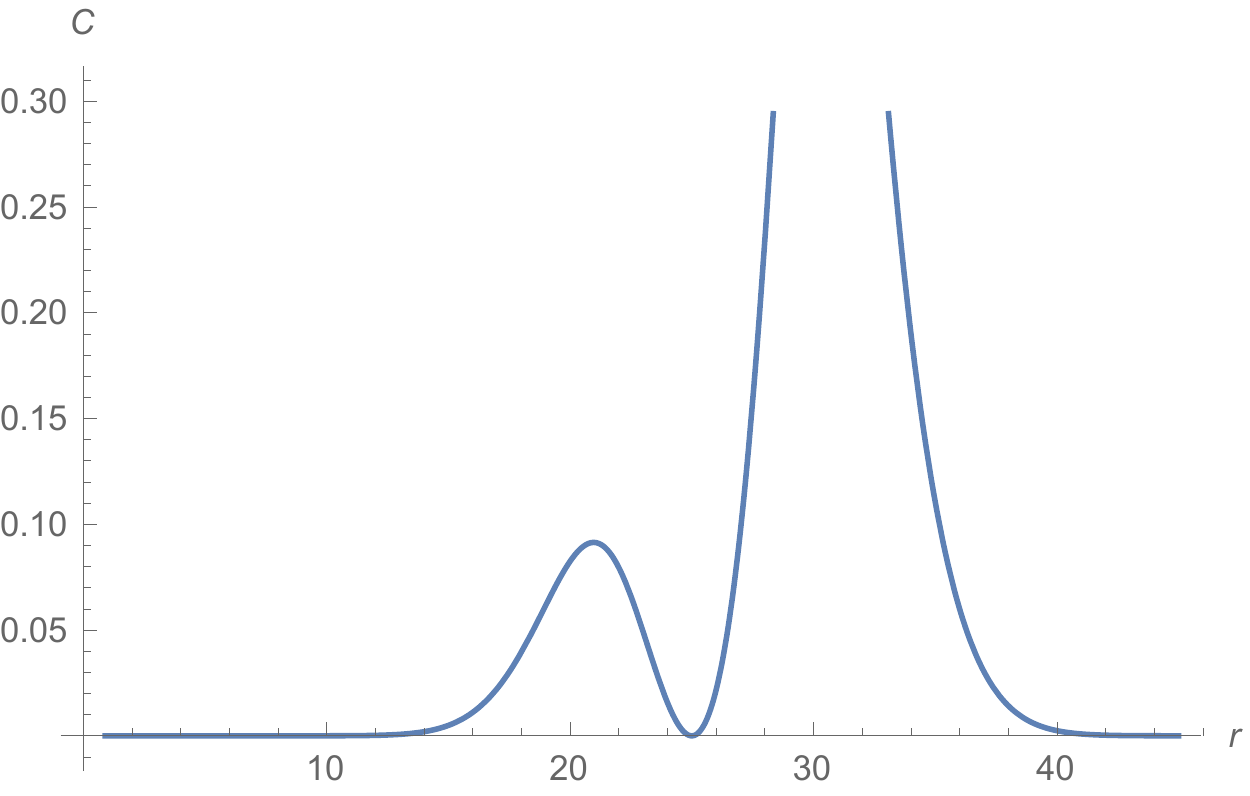}}
\subfigure[\ ] {
\includegraphics[scale=0.35]{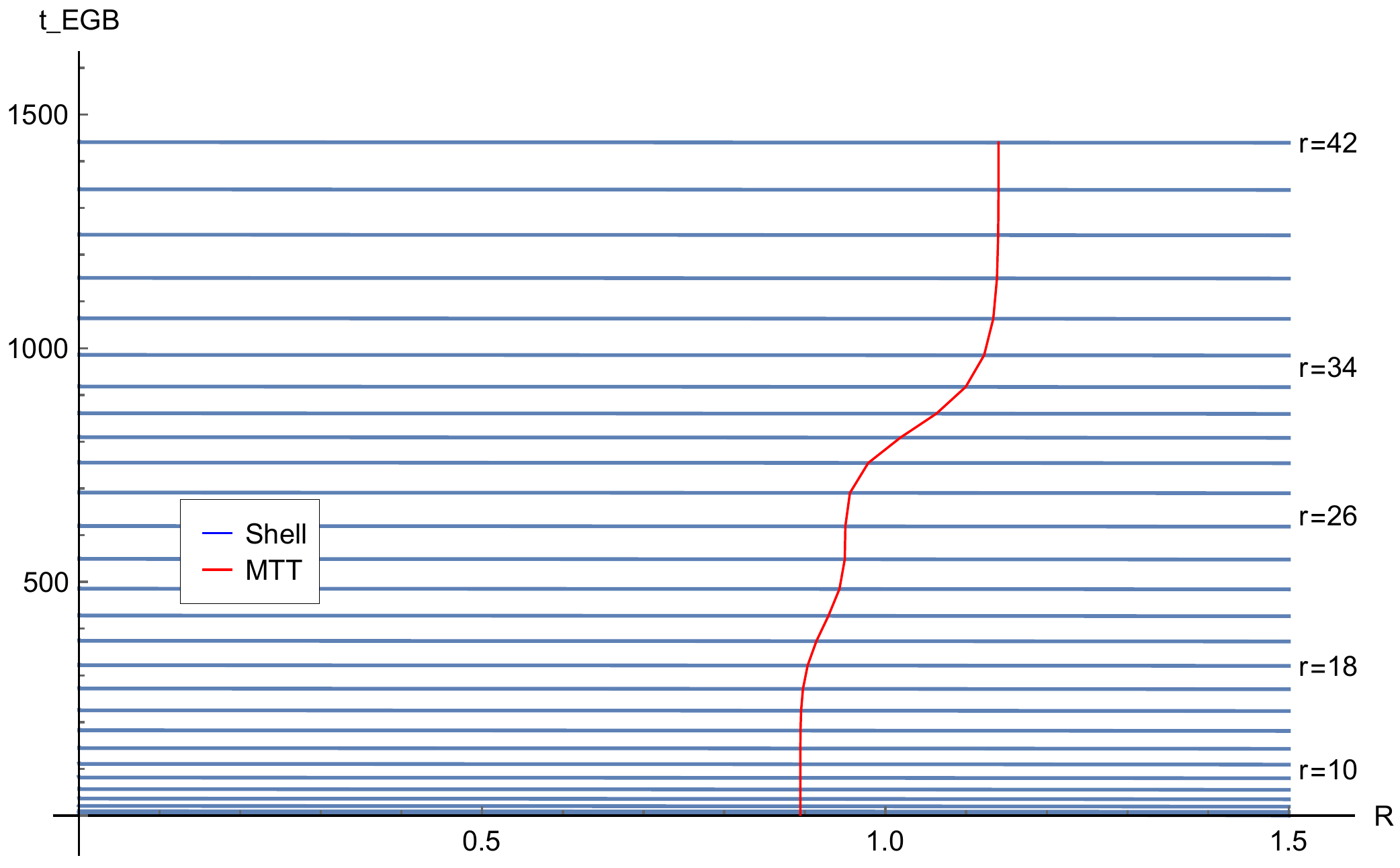}}
\caption{\label{fig:Maxkg0_example4EGB} The graphs show the (a) values of $C$,
and (b) formation of MTT along with the shells which fall consecutively 
on a black hole. The value of $C$ remains positive and large, and for that
reason it is not plotted here. As a consequence MTT remains spacelike.}
\end{figure}
%
The graphs corresponding to this case is given in figure \eqref{fig:Maxkg0_example4EGB}.
Note that these graphs have a similar behaviour to those in figure \eqref{fig:Maxk0_example4EGB},
except that the times for formation of MTTs have changed.


%
\item Let us again consider the density profile given by 
eqn. \eqref{timelike_rho}, given in figure \eqref{fig:Timelikek0_example2EGB}(a).
%
\begin{figure}[t!]
\centering
\subfigure[\ ] {
\includegraphics[scale=0.4]{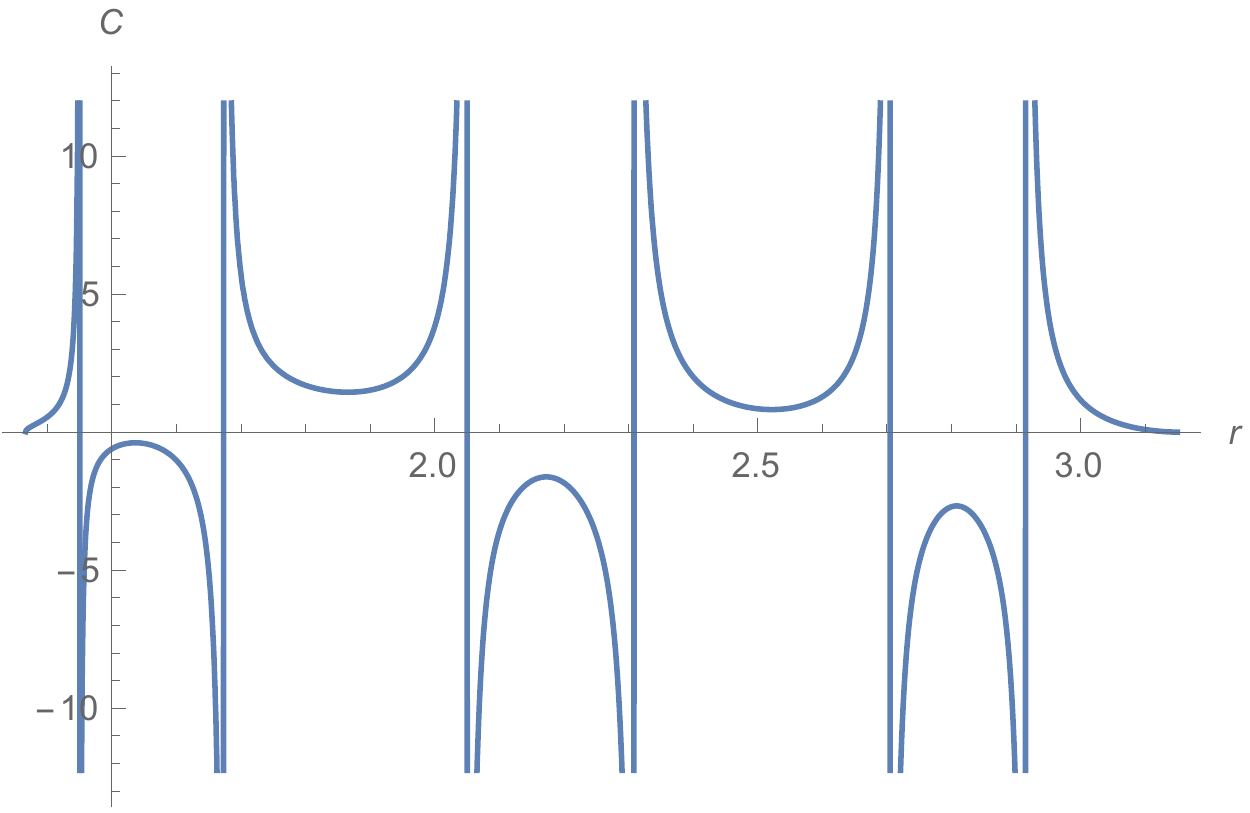}}
\subfigure[\ ] {
\includegraphics[scale=0.35]{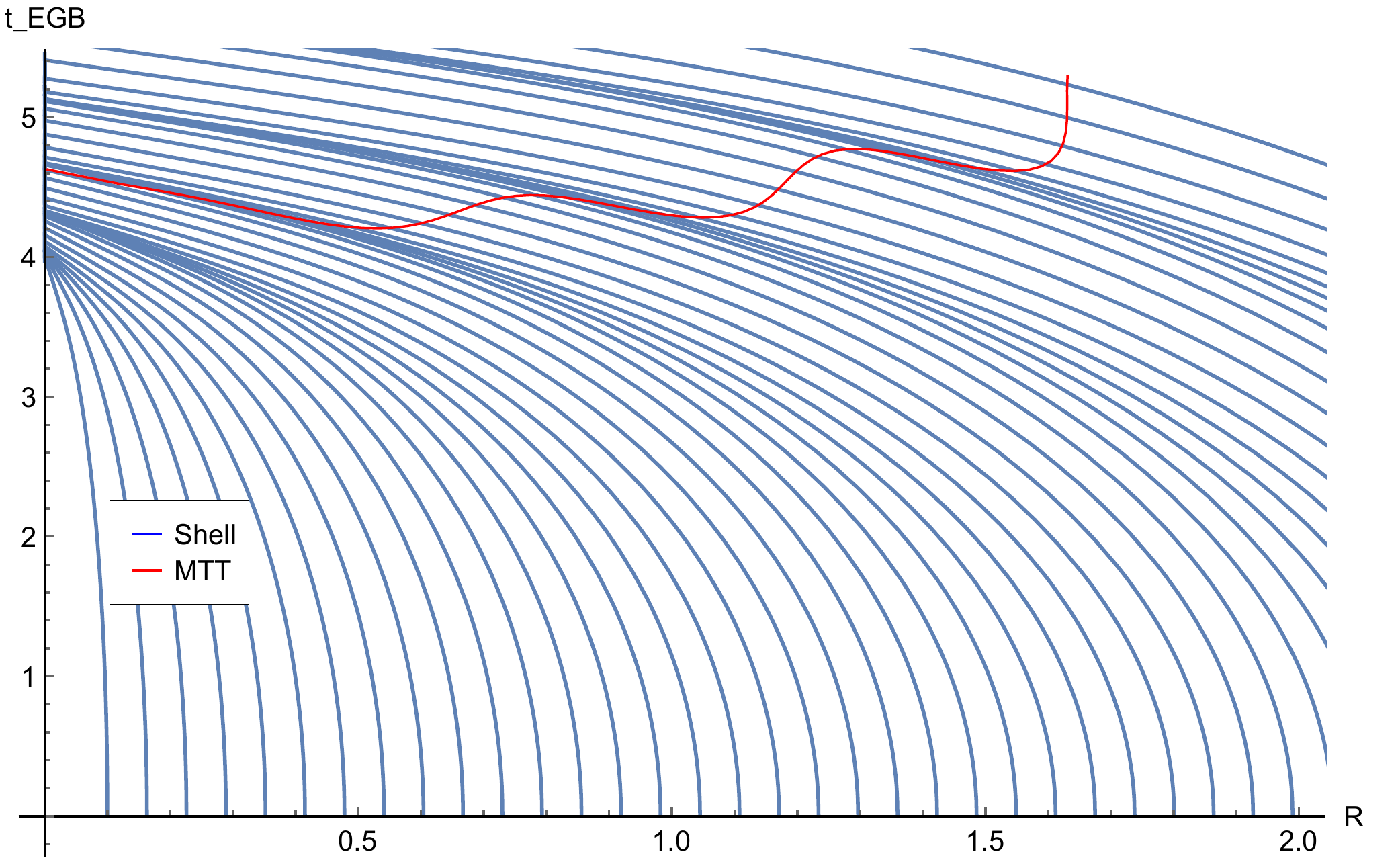}}
\caption{\label{fig:Timelikekg0_example2EGB} The graphs show the (a) density distribution, (b) values of $C$,
and (c) formation of MTT along with the shells for the bounded collapse of the density profile discussed
in eqn. \eqref{timelike_rho}. The MTT is quite complicated and goes through various modulations.}
\end{figure}
%
The behaviour is of the MTTs and the shells, for the
bounded collapse as given in figure \eqref{fig:Timelikekg0_example2EGB} is similar to the 
graphs in figure \eqref{fig:Timelikek0_example2EGB},
with the exception that the time of formation of MTTs, and as well as those of the shells reaching the singularity
has changed. 

\end{enumerate}

Similar study may be carried out for more complicated matter profiles 
and other matter sources. These studies can be made using the techniques developed above.

\section{Discussions}\label{sec6}

This paper deals with the study of gravitational collapse 
in EGB gravity in $5$- dimensions. The Gauss- Bonnet modification of the
Einstein gravity changes the geometry of the spacetime, and the structure of the horizon and singularity quite drastically. 
We developed techniques to analyse these effects in the
phenomena of gravitational collapse in this theory. In this context, several 
questions arise naturally regarding the process of the collapse phenomenon itself as well as the 
outcome of gravitational collapse of matter. To understand 
these details, we have, in this paper, developed a set of analytical and numerical techniques 
to locate spherical marginally trapped surfaces in the spacetime, when the collapse is
in progress. We locate these MTTs for a large class of matter profiles and initial velocity profiles.
This study helps us to address several questions regarding gravitational
collapse in the EGB theory:
%
\vspace{0.2cm}

(i)\emph{ Role of the GB term and the coupling constant $\lambda$:} The GB term introduces 
several changes in the equation of motion of the gravitational field. The most drastic is the 
change in the form of the mass function $F(r,t)$ given in eqn. \eqref{1eq5EGB}. In fact, this equation shows 
that the GB term leads to quadratic effects involving $\dot{R}(r,t)$ and $R^{\prime}(r,t)$.
As a result of this quadratic contribution of $\dot{R}$, the equation of motion
of the radius of the dust cloud is altered significantly, see eqn. \eqref{rdoteqn}. 
Naturally, this change in the equation of motion of the spherically symmetric 
matter configuration implies that the collapsing matter spheres will get trapped
at different times. A direct reflection of this fact is in the expressions for
the expansion of the outward and the inward 
null normals $\theta_{(\ell)}$ and $\theta_{(n)}$ in 
eqns. \eqref{null_normals_exp} and \eqref{null_normals_exp_2} . It follows 
as a direct result of \eqref{null_normals_exp_2} that the equation defining a marginally 
trapped surface is dependent on the GB coupling constant $\lambda$,
see eqn. \eqref{trapped_surface_eqn}. The marginally 
trapped surface (MTS) forms at $R_{M}(r,t)=F(r,t)^{1/2}$ in the $5$- dimensional Einstein 
theory, whereas it forms at $\sqrt{F(r,t)-2\lambda}$ in the EGB theory. In this
paper, we have kept the value of $\lambda=0.1$, and so, the equation for MTS, 
eqn. \eqref{trapped_surface_eqn} implies that real values of $R_{M}(r,t)$ is only 
possible only if sufficient number of shells have fallen in so that the cloud if massive enough
to overcome the effect of the GB coupling constant $\lambda$. This effect on 
the formation of a MTS and the MTT is directly visible in 
the graphs in fig. \eqref{fig:ltbk0_example2EGB}, fig. \eqref{fig:Gaussk0_example2EGB} as well as in
the fig. \eqref{fig:Gausskg0_example2EGB}. The coupling constant 
results in the delay in the formation of MTT, and as can be noticed
from these figures, begins to form quite later than the formation of central singularity
due collapsing shells. This effect is not visible in fig. \eqref{fig:Maxk0_example4EGB},
since the system already has a spacetime singularity, and so, this initial black hole
horizon censors all the singularities arising out of shell collapse. 

\begin{figure}[t!]
\centering
\includegraphics[width=0.6\linewidth]{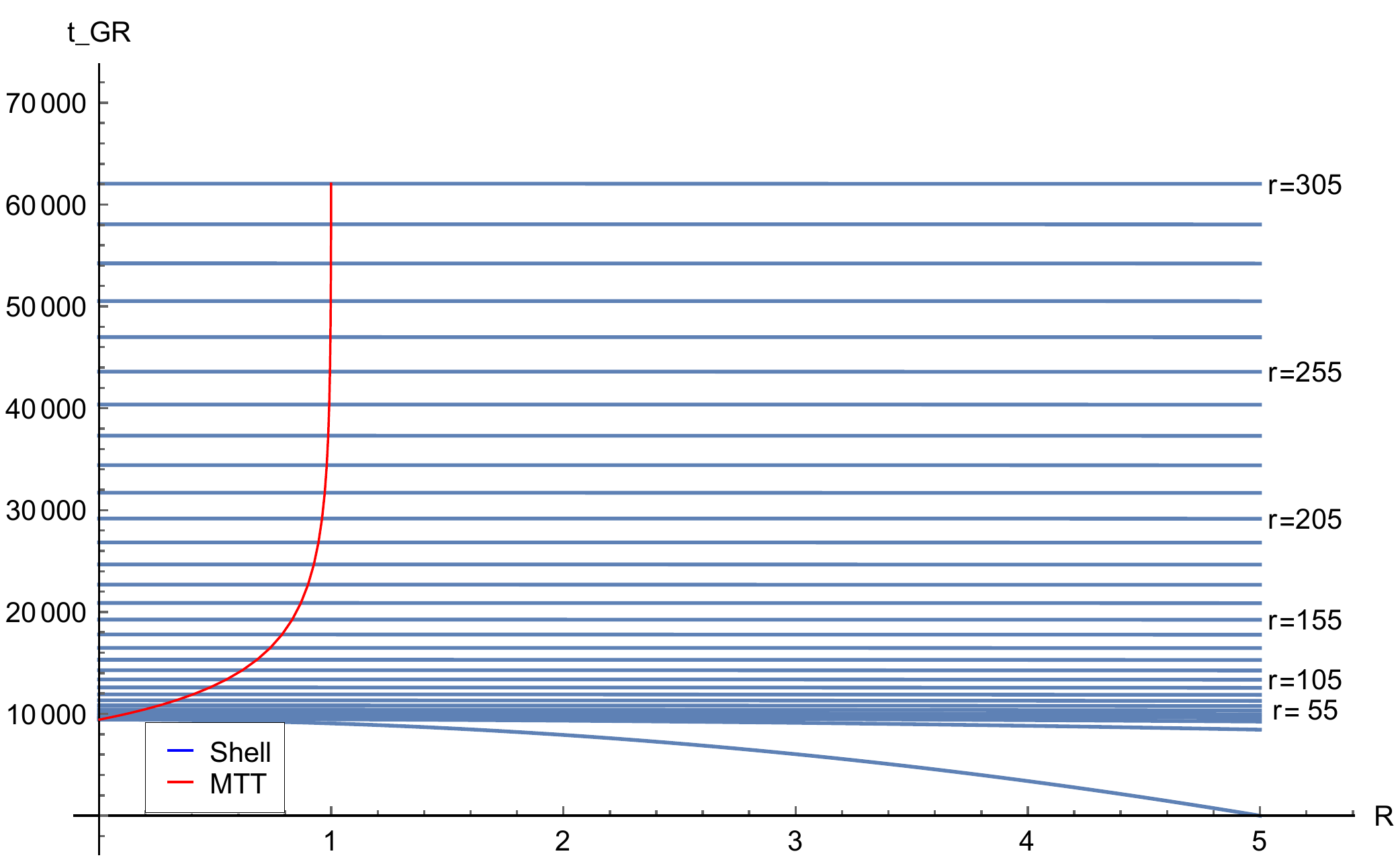}
\caption{The graphs show formation of MTT along with the shells for marginally bound collapse
of pressureless matter with a Gaussian density profile of  eqn. \eqref{rho_gaussian}, in $5$- dimensional GR.}
\label{fig:MTT_5D}
\end{figure}

It is also instructive to compare this same study of MTTs for 
the Gaussian profile in eqn. \eqref{rho_gaussian}
in the  $5$- dimensional Einstein 
theory. As expected, the MTT begins just as the first shells start to collapse and 
the MTT equilibriates at $R=1$, since the total mass of the profile is unity, and
the MTT is $R_{M}(r,t)=F(r,t)^{1/2}$. This is given in fig. \eqref{fig:MTT_5D}.

\vspace{0.2cm}
(ii) \emph{Nature of the central singularity:} Since many of these configurations
lead to shell collapsing \emph{naked} singularities due to gravitational collapse of the initial shells, and that
MTTs do not cover them, it 
becomes essential to characterise them, and make a clear classification. We have explicitly verified 
that, in each of the cases where the central singularity is naked initially, satisfy the following
relation: \emph{The weak cosmic censorship is violated for each of these collapse processes
until the mass function $F(r)\, > \, 2\lambda $} (see also \cite{Maeda:2006pm}). 
The fact that the curvature strength of 
the singularity is a weak is obtained as follows: Note that 
the singularity is defined to be strong if 
the spacetime volume contained within Jacobi vector fields is reduced to zero at the singularity. 
The singularity is weak otherwise. According to the standard characterizations
of singularities in $4$- dimensions \cite{joshi}, a sufficient condition  for a 
strong singularity is that at least one causal geodesic $t^{\mu}$,
with affine parameter $v$
must satisfy the following condition:
\begin{equation}
\lim_{v\rightarrow v_{0}}\, (v\rightarrow v_{0})^{2}\, R_{\mu\nu}t^{\mu}t^{\nu}>0
\end{equation}
For our spacetime, and a radial timelike vector field, a 
simple calculation shows that $\lim_{v\rightarrow v_{0}}\, (v\rightarrow v_{0})^{2}\, R_{\mu\nu}t^{\mu}t^{\nu}=0$. 
Here too, the role
of the Gauss- Bonnet coupling becomes crucial, and plays an important role in weakening the singularity.
So, although the singularities are naked at the beginning of the collapse process, the singularity
is harmless since they are weakly naked.
%

\vspace{0.2cm}
(iii)\emph{Are MTT  true black hole boundaries? }
The actual extent of a black hole region is a matter of great debate.
Over the years, global as well as quasilocal considerations
have led to several formulations of horizon. Out of them,
event horizon, Killing horizons have been quite useful to study physical phenomena of black holes.
The quasilocal formulations based on trapped surfaces, and in particular
the definitions of trapping horizons and MTTs  \cite{Hayward:1993wb, Ashtekar:2005ez}
have been extensively used to prove classical and quantum 
laws of black hole dynamics. Although, it must also be pointed out that the formulation
of MTT as a black hole boundary may need modifications, in particular in respect to the
conditions on $\theta_{(n)}$, they may be quite useful for this purpose.
However, the main issue lies in locating the non- spherically symmetric MTTs as well,
and in the context of $4$- dimensional spacetimes, they are yet to be 
completely specified \cite{Bengtsson:2008jr,Bengtsson:2010tj,Bengtsson:2013hla}. 
Furthermore, for some spacetimes, the black hole boundary is identical with the event 
horizon \cite{Eardley:1997hk, BenDov:2006vw}.
Our study using spherical MTTs in $5$- dimensions show that they may indeed be used as a boundary of a black hole 
region, although a non- spherical MTTs and their location is equally important to be understood in this
context. We must also point out that our study needs to be extended for more general matter fields and geometries,
so that such questions may be included in our discussions.

To conclude, we have explicitly shown, with a wide range of examples, that the nature of trapped surface,
its formation and time development, is intimately related to the initial velocity and the initial density profile of the matter fields.
Additionally, due to the presence of the EGB coupling constant $\lambda$, the formation of MTT gets delayed further,
depending on the amount of matter a particular matter shell encloses within its boundaries. All these effects
have been conclusively demonstrated through the examples considered in the main part of the paper. We must however
admit that a full understanding of these phenomenon of gravitational collapse and the censorship conjecture
shall require the methods of non-spherical gravitational collapse.


\section*{Acknowledgements}
The author AC is supported through the DAE-BRNS project $58/14/25/2019$-BRNS,
and by the DST-MATRICS scheme of government of India through their grant MTR-$/2019/000916$. 
AG acknowledges the support through grants from the NSF of China with Grant No: 11947301
and Fundamental Research Funds for Central universities under grant no. WK2030000036.


\section{Appendix}

\subsection{Expressions for curvature using matter variables}
\label{matter_appendix}
In the following, we collect the expressions of the various
curvature components for the metric \eqref{1eq1EGB}. These components have been used in the
main part of the paper to determine the evolution of MTT, and in determining
the signature of the MTT in eqn. \eqref{OSDCEGB}.
The quantities like the Ricci scalar ($R_{s}$), Ricci tensors and the Riemann tensors in terms 
of the energy density, radial and tangential pressure and mass function.

First, the Riemann tensors are obtained using the metric functions and the matter variables:
\begin{eqnarray*}
R_{\theta\phi\theta\phi}&=&F(r,t) \sin^2\theta\, , \qquad
R_{\theta\psi\theta\psi}=\sin^2{\theta} R_{\theta\phi\theta\phi}\, ,
\qquad R_{\phi\psi\phi\psi}=\sin^2{\theta}\sin^2 {\phi}\, R_{\theta\phi\theta\phi}, \nonumber\\
R_{t\phi t\phi}&=&\sin^2{\theta} R_{t\theta t\theta}\, ,\qquad 
R_{t\psi t\psi}=\sin^2{\theta}\sin^2{\phi}\, R_{t\theta t\theta}, \qquad R_{t\theta r \theta}=0 \nonumber\\
R_{r\phi r\phi}&=&\sin^2{\theta}\, R_{r\theta r\theta}\, , \qquad
R_{r\psi r\psi}=\sin^2{\theta}\sin^2{\phi}\, R_{r\theta r\theta}, \nonumber\\
R_{t\theta t\theta}&=&-(1/2)e^{2\alpha}\frac{R}{\dot{R}}\,\frac{d}{dt} \left[\frac{F(r,t)}{R^{2}(r,t)}-1\right]\, \nonumber\\
R_{r\theta r\theta}&=&(1/2)\,e^{2\beta}\frac{R}{R^{\prime}}\frac{d}{dr}
\left[\frac{F(t,r)}{R^{2}(t,r)}-1\right]\, , \nonumber\\
R_{rtrt}&=&[p_{t}-(2/3)\left(\rho+p_{r}\right)-(3F/R^{\,4})]\, e^{2(\alpha+\beta)}.\nonumber\\
\end{eqnarray*}
The Ricci tensors are obtained similarly using the metric in eqn. \eqref{1eq1EGB}.
\begin{eqnarray*}
R_{tt}&=& (-2\rho/3 + p_{r}/3 +p_{t})e^{2\alpha},\qquad
R_{rr}=(-\rho/3 +2p_{r}/3-p_{t})e^{2\beta},\label{Rrr} \nonumber\\
R_{\theta\theta}&=&-(R^{2}/3)\left(\rho+p_{r}\right),\qquad 
R_{\phi\phi}=\sin^2 \theta\, R_{\theta\theta},\\ 
R_{\psi\psi}&=&\sin^2{\theta}\sin ^2{\phi}\, R_{\theta\theta}, \qquad R_{rt}=0,
\end{eqnarray*}
The Ricci scalar is given by $R_{s}=-(2/3)\left(\rho+p_{r}\right)-2 p_{t}$. Using these expressions, 
and the expressions for null normals in eqns. \eqref{null_normals_exp} and 
\eqref{null_normals_exp_2}, it can be shown easily that:
\begin{eqnarray*}
H_{ab}\ell^{a}\ell^{b}
&=&
2\left[\frac{{6F}\left(\rho+p_{r}\right)}{\left(F-2\lambda\right)^2}
+2p_{\theta}^2-4p_{r}p_{\theta}-\frac{2}{3}p_{\theta}\left(\rho+p_{r} \right) \right],\\
H_{ab}\ell^{a}n^{b}
&=&2\left[4p_{\theta}\left(p_{\theta}+\frac{2}{9}\rho-\frac{4}{3}p_{r}\right)-\frac{2}{\left(F-2\lambda\right)^2}\{6F p_{t}+\left(F+4\lambda\right)\left(\rho-p_{r}\right)\}\right.\nonumber\\
&& ~~~~ -\left.6\left\{p_{t}+\frac{2}{3}\left(\rho-p_{r}\right)-\frac{3F}{\left(F-2\lambda\right)^2} \right\}^{2}
+\frac{16}{9}\left(\rho^2+p_{r}^2\right)-72\frac{\lambda^{2}}{\left(F-2\lambda\right)^{4}}\right].\label{Hlanb}
\end{eqnarray*}
We can also similarly determine an expression for $L_{GB}$ in 
terms of matter variables and the mass function.  
\begin{eqnarray*}
L_{GB}&=&R^{2}-4R_{tt}R^{tt}-4R_{rr}R^{rr}-12R_{\theta\theta}R^{\theta\theta}
+6R_{trtr}R^{trtr}+18R_{t\theta t\theta}R^{t\theta t\theta}
+18R_{r\theta r\theta}R^{r\theta r\theta}+18R_{\theta \phi\theta\phi}R^{\theta \phi\theta\phi}\nonumber\\
&=&\left[\frac{2}{3}\left(\rho-p_{r} \right)-2p_{t} \right]^2 
+ 18\left[ \frac{F^{2}+32\lambda^2}{\left(F-2\lambda\right)^{4}}\right]+6\left[p_{t}+\frac{2}{3}\left(\rho-p_{r}\right)-\frac{3F}{\left(F-2\lambda\right)^2} \right]^{2}\nonumber\\
&&~~~~~~~~ ~~~~~~~~~~~~~ ~~~~~~~~~~~
-\frac{12}{9}\left(\rho-p_{r} \right)^2-4 \left[\frac{2}{3}\left(\rho+p_{r} \right)+p_{t} \right]^2 -4 \left[\frac{2}{3}\left(\rho+p_{r} \right)-p_{t} \right]^2 .\label{LGBlanb}
\end{eqnarray*}



\subsection{Three- surface geometry}
\label{geometry_appendix}
The subspace in our problem
is a three dimensional sphere. To understand the geometry
of this subspace, we shall present a general formulation of subspaces.
Let $(\mathcal{M},\, g_{\mu\nu}, \nabla_{\mu})$ be a $5$- dimensional time- oriented spacetime
with a metric compatible covariant derivative $\nabla_{\mu}\,g_{\nu\lambda}=0$. Let us 
assume that $\mathcal{S}$ be a closed, orientable, spacelike $3$- surface embedded in $\mathcal{M}$.
Let us denote the two future pointing null vectors by $\ell^{\mu}$ (outward pointing) and
$n^{\mu}$ (inward pointing), such that $\ell\cdot n=-1$.

The induced metric $h_{ab}$ on the $3$- surface $\mathcal{S}$ is given by:
\begin{equation}
h_{ab}=e^{\mu}{}_{a}e^{\nu}{}_{b}\, g_{\mu \nu},
\end{equation}
where $e^{\mu}{}_{a}$ denotes the pullback map, and $a, b, \dots$ indicate indices on $\mathcal{S}$.
The functions $e^{\mu}{}_{a}$ are orthogonal to  $\ell^{\mu}$ and $n^{\mu}$. This implies 
that the pushforward of the inverse two- metric $h^{ab}$ is given by:
\begin{equation}
g^{\mu\nu}=e^{\mu}{}_{a}e^{\nu}{}_{b}\, h^{ab}- \ell^{\,\mu}n^{\nu}- \ell^{\,\nu}n^{\mu}.
\end{equation}
The second important quantities of importance are the extrinsic curvatures. This is vector
on the normal bundle $N(\mathcal{S})$ of $\mathcal{S}$, and it has two components.
\begin{equation}
k^{(\ell)}{}_{ab}=e^{\mu}{}_{a}e^{\nu}{}_{b}\, \nabla_{\mu}\, \ell_{\nu}, ~~~~~~
k^{(n)}{}_{ab}=e^{\mu}{}_{a}e^{\nu}{}_{b}\, \nabla_{\mu}\,n_{\nu},
\end{equation}
where the extrinsic curvature itself may be written as:
\begin{equation}
k^{\,\mu}{}_{ab}=k^{(n)}{}_{ab}\,\ell^{\mu}+k^{(\ell)}{}_{ab}\, n^{\mu}.
\end{equation}
The Riemann tensor on $\mathcal{M}$ and on $\mathcal{S}$ are given respectively by:
\begin{eqnarray}
(\nabla_{\mu}\nabla_{\nu}- \nabla_{\nu}\nabla_{\mu})Z_{\lambda}&=&R_{\mu\nu\lambda\sigma}\,Z^{\sigma}\\
(\mcD_{a}\mcD_{b}- \mcD_{b}\mcD_{a})z_{c}&=&\mathcal{R}_{abcd}\,z^{d},
\end{eqnarray}
where $\mcD$ is the metric compatible derivative operator on $\mcS$, so that $\mcD_{a}h_{bc}=0$.
The Gauss equation for the spacetime and submanifold gives the following equation:
\begin{equation}
e^{\mu}{}_{a}e^{\nu}{}_{b}e^{\lambda}{}_{c}e^{\sigma}{}_{d}\, R_{\mu\nu\lambda\sigma}
=\mcR_{abcd}-(k^{(\ell)}{}_{ac}k^{(n)}{}_{bd}+k^{(n)}{}_{ac}k^{(\ell)}{}_{bd})+
(k^{(\ell)}{}_{ad}k^{(n)}{}_{bc}+k^{(n)}{}_{ad}k^{(\ell)}{}_{bc}),
\end{equation}
and the Codazzi equations may be written in the following forms corresponding to each of 
the two normals:
\begin{eqnarray}
e^{\mu}{}_{a}e^{\nu}{}_{b}e^{\lambda}{}_{c}\,\ell^{\sigma}\, R_{\mu\nu\lambda\sigma}
&=&(D_{b}-\omega_{b})k^{(\ell)}{}_{ac} - (D_{a}-\omega_{a})k^{(\ell)}{}_{bc}\\
e^{\mu}{}_{a}e^{\nu}{}_{b}e^{\lambda}{}_{c}\, n^{\sigma}\, R_{\mu\nu\lambda\sigma}
&=&(D_{b}-\omega_{b})k^{(n)}{}_{ac} - (D_{a}-\omega_{a})k^{(n)}{}_{bc},
\end{eqnarray}
where $\omega_{\underleftarrow{a}}\equiv e^{\mu}{}_{a}\,\omega_{\mu}$ is 
the pullback of the connection on 
the normal bundle $\mathcal{N}(\mcS)$, and is defined using the equation for the
Shape operator to get: 
$\omega_{\underleftarrow{a}}=-n_{\sigma}\,e^{\lambda}{}_{a}\,\nabla_{\lambda}\ell^{\sigma}$.

The variation of the submanifold $\mcS$ in the normal direction $N^{\mu}=A\ell^{\mu}-Bn^{\mu}$, $A$ 
and $B$ being constants, is given by the variation of the abovementioned spacetime variables.
The variation in the induced metric is:
\begin{equation}
\nabla_{N}h_{ab}=2Ak^{(\ell)}_{ab}-2Bk^{(n)}_{ab}
\end{equation}
whereas, the variation of the area element $\sqrt{h}=\sqrt{\det h_{ab} }$ is given by:
\begin{equation}
\nabla_{N}\sqrt{h}=(1/2)\sqrt{h}\, h^{ab}\,\nabla_{N}h_{ab}=(A\theta_{(\ell)}-B\theta_{(n)})\sqrt{h}.
\end{equation}
The extrinsic curvatures are written in terms of the expansion scalar and the shear tensors
of the two null normals:
\begin{equation}
k^{(\ell)}{}_{ab}=\frac{1}{(D-2)}\,\theta_{(\ell)}h_{ab}+\sigma^{(\ell)}_{(ab)},~~~~
k^{(n)}{}_{ab}=\frac{1}{(D-2)}\, \theta_{(n)}h_{ab}+\sigma^{(n)}_{(ab)} 
\end{equation}
where the expansion scalar and the shear tensors are defined as:
\begin{eqnarray}
\theta_{(\ell)}&=&\nabla_{\mu}\ell^{\mu}-\kappa_{(\ell)}\\
\sigma^{(\ell)}_{ab}&=&\left[e^{\mu}{}_{a}e^{\nu}{}_{b}-
\frac{h_{ab}}{(D-2)}\, g^{\mu\nu}\right]\nabla_{\mu}\ell_{\nu}
+\kappa_{(\ell)}\,h_{ab},
\end{eqnarray}
where $\kappa_{(\ell)}=-n_{\nu}\ell^{\mu}\,\nabla_{\mu}\ell^{\,\nu}$
is the measure of affinity of the null normal. 
These equations for the other null- normal $n^{\mu}$
is obtained by $\ell^{\mu}\leftrightarrow n^{\mu}$. 

Let us now consider how the foliation is evolved along $N^{\mu}$. Since $\ell_{\mu}$ and $n_{\mu}$
are normal to $\mcS$, their pullback on $\mcS$ vanish. 
Thus, $e^{\mu}{}_{a}\ell_{\mu}\equiv \ell_{\underleftarrow{a}}=0$, and also the same is 
true for $n_{\underleftarrow{a}}$. This foliation is assumed to be preserved in the evolution under 
$N^{\mu}$, so that $(\lie_{N}\ell)_{\underleftarrow{a}}=0$, and $(\lie_{N}n)_{\underleftarrow{a}}=0$
is assumed to hold true. These equations imply that:
\begin{eqnarray}
N^{\mu}\nabla_{\mu}\ell_{\underleftarrow{a}}&=&\kappa_{(N)}\ell_{\underleftarrow{a}}
-(D_{\underleftarrow{a}}-\omega_{\underleftarrow{a}})B,\\
N^{\mu}\nabla_{\mu}n_{\underleftarrow{a}}&=&-\kappa_{(N)}n_{\underleftarrow{a}}
+(D_{\underleftarrow{a}}+\omega_{\underleftarrow{a}})B,
\end{eqnarray}
where $\kappa_{(N)}=-n_{\mu}N^{\nu}\nabla_{\nu}\ell^{\mu}$ is called the surface gravity
corresponding to the vector field $N^{\mu}$. A direct calculation leads to the following results on 
the variation of $\theta_{(\ell)}$ \cite{Booth:2006bn}
\begin{eqnarray}
\nabla_{N}\theta_{(\ell)}-\kappa_{N}\theta_{(\ell)}&=&-d^{2}B+2\omega^{\mu}d_{\mu}B- B[\omega^{\mu}\omega_{\mu}-d_{\mu}\omega^{\mu}
-(R/2)-G_{\mu\nu}\ell^{\mu}n^{\nu}-\theta_{(\ell)}\theta_{n}\nonumber]\\
&& ~~~~~~~~~~~~~~~~~~~~~~~~~~~~~~~~~ -A[\sigma_{(\ell)}^{2}+G_{\mu\nu}\ell^{\mu}\ell^{\nu} +(1/2)\theta_{(\ell)}^{2}].
\end{eqnarray}
%


\subsection{Matching conditions at shell boundary}
\label{matching_appendix}
In the following, we present the junction condition of a LTB metric,
formed due to collapse of a spherically symmetric matter configuration,
with the spherically symmetric metric due to a body of mass $M$.
The interior LTB metric of the spacetime $\mathcal{M^{-}}$ is given by eqn. \eqref{int_metric}:
\begin{equation}
ds_{-}^{2}=-dt^2 + \frac{R^{\, \prime \, 2}}{1-k(r)}\,dr^2 + R(r,t)^{\, 2} \,d\Omega_{3},
\end{equation}
where $d\Omega_{3}$ is the metric of an
unit round $3$-sphere, and $R(t,r)$ is obtained from 
the equation \eqref{1eq5EGB}.
The metric 
of the external spacetime $\mathcal{M_{+}}$ is the 
Boulware- Deser- Wheeler solution \cite{Boulware:1985wk, Wheeler:1985nh, Wheeler:1985qd,Torii:2005xu},
which for $5$- dimensions is given by:
\begin{equation}
ds_{+}^{2}=-F(\bar{R})\, dT^{2} + F(\bar{R})^{-1}\, d\bar{R}^{2}+\bar{R}^{\, 2} \,d\Omega_{3}, 
\end{equation}
where $T$ and $\bar{R}$ are the time and radial coordinates in $\mathcal{M}_{+}$, and the metric function 
$F(\bar{R})$ is:
\begin{equation}
F(\bar{R})=1+\frac{\bar{R}^{2}}{4\lambda}\left[1 \mp \sqrt{1+\frac{8\lambda M}{\bar{R}^{4}}}\right]
\end{equation}
gives the external vacuum solution for a spherical body of mass $M$ when the $-$ve sign is chosen.

The matching is to be carried out at the timelike hypersurface $\Sigma$ given by $r_{b}$.
Let us denote the coordinates on this surface $\Sigma$ to be $(\tau, \theta, \phi, \psi)$.  
From $\mathcal{M^{-}}$, we can write down the surface $\Sigma$ as
$f_{-}(r,t)=r-r_{b}=0$, and hence, the induced metric on $\Sigma$ is
\begin{eqnarray}
ds^{2}_{-}=-d\tau^{2}+ r_{b}^{2}\, d \Omega_{3}\,.
\label{M_1}
\end{eqnarray}
From the point of view of the exterior spacetime, the hypersurface
may be described by $r=\bar{R}_{\Sigma}(\tau)$ and $t=T_{\Sigma}(\tau)$, with no change in 
the angular variables. 
The line element of the hypersurface is then given by
\begin{equation}
ds^{2}_{+}=-\left[F(\bar{R}_{\Sigma})\dot{T}_{\Sigma}^2-F(\bar{R}_{\Sigma})^{-1}\dot{\bar{R}}_{\Sigma}^2\right]d\tau^2 
+ \bar{R}_{\Sigma}(\tau)^2 d \Omega_{3}\,,
\label{m11+}
\end{equation}
where the dots imply derivative with respect to $\tau$.

The induced metric in equations in \eqref{M_1} and \eqref{m11+} must have matched metric functions.
This implies that:
\begin{eqnarray}\label{metric_matching}
F(\bar{R}_{\Sigma})\dot{T}_{\Sigma}^2-F(\bar{R}_{\Sigma})^{-1}\dot{\bar{R}}_{\Sigma}^2=1
\end{eqnarray}

Now, let $u^{\mu}$ and $n^{\mu}$ denote the velocity of the matter variables and the 
normal to the $\Sigma$ respectively. They must satisfy the conditions 
$u^{\mu}u_{\mu}=-1$, $n^{\mu}n_{\mu}=1$, whereas, $u^{\mu}n_{\mu}=0$. From the 
interior spacetime, the expressions of these vectors is easily obtained:
\begin{equation}
u^{\mu}=\delta^{\mu}_{0}\equiv (\partial_{\tau})^{\mu}, 
\, \qquad  \, n_{\mu}=\frac{R^{\, \prime }}{\sqrt{1-k(r)}}\,(dr)_{\mu} .
\end{equation}
From the exterior spacetime, these vectors are also obtained similarly to give:
\begin{equation}
u^{\mu}=\dot{T}_{\Sigma}\,\,(\partial_{\tau})^{\mu} + \dot{\bar{R}}_{\Sigma}\,(\partial_{r})^{\mu} , 
\, \qquad  \, n_{\mu}=-\dot{\bar{R}}_{\Sigma}\,(d\tau)_{\mu} 
+ \dot{T}_{\Sigma}\,\,(d{r})_{\mu} .
\end{equation}

The extrinsic curvatures are easily determined from these normals for the exterior as well the interior
spacetimes:
\begin{eqnarray}
&K^{-}_{\tau\tau}=0, \,\, & \qquad K^{-}_{\theta\theta}= \bar{R}_{\Sigma}\,\sqrt{1-k(r_{b})}\\
 &K^{+}_{\tau\tau}=\dot{\bar{R}}_{\Sigma}^{-1}[\dot{F}(\bar{R}_{\Sigma})\, \dot{T}_{\Sigma}
+{F}(\bar{R}_{\Sigma})\, \ddot{T}_{\Sigma}\, ] \, 
 &\qquad K^{+}_{\theta\theta}= \bar{R}_{\Sigma}\,{F}(\bar{R}_{\Sigma})\, \dot{T}_{\Sigma}.
\end{eqnarray}
The $K_{\theta\theta}$ equations imply the following relation:
\begin{equation}
\frac{dT_{\Sigma}}{d\tau}=\frac{\sqrt{1-k(r_{b})}}{F(\bar{R}_{\Sigma})},
\end{equation}
whereas the equation \eqref{metric_matching} gives the following equation for the function 
$\bar{R}_{\Sigma}$ :
\begin{equation}
\frac{d\bar{R}_{\Sigma}}{d\tau}=[1-k(r_{b})-F(\bar{R}_{\Sigma})]^{1/2}.
\end{equation}
This implies that the following relation hold good:
\begin{equation}
\left(\frac{d\bar{R}_{\Sigma}}{d\tau}\right)^{2}=-k(r_{b})
+\frac{\bar{R}_{\Sigma}^{2}}{4\lambda}\left[1 \mp \sqrt{1+\frac{8\lambda M}{\bar{R}_{\Sigma}^{4}}}\right].
\end{equation}
A simple comparison with equation \eqref{EQMEGB} implies that the condition $M=F(r_{b})$
must be satisfied at the boundary.

%
 

\end{document}